\documentclass[10pt]{article}
\usepackage{amsmath}
\usepackage{amssymb}
\usepackage{color}
\usepackage{graphicx}
\usepackage{array, multirow}
\usepackage{MnSymbol}
\usepackage{url}
\newcolumntype{C}[1]{>{\centering\let\newline\\\arraybackslash\hspace{0pt}}m{#1}}

\title{Templates and subtemplates of R\"ossler attractors from a bifurcation diagram}

\author{Martin Rosalie\\
Univ. Bordeaux, LaBRI, UMR5800, F-33400, Talence, France\\
\url{martin.rosalie@labri.fr}
}

\begin{document}

\maketitle

\vspace{10pt}

\begin{abstract}
    We study the bifurcation diagram of the R\"ossler system. It displays the
    various dynamical regimes of the system (stable or chaotic) when a
    parameter is varied. We choose a diagram that exhibits coexisting
    attractors and banded chaos. We use the topological characterization
    method to study these attractors. Then, we details how the templates of
    these attractors are subtemplates of a unique template. Our main result is
    that only one template describe the topological structure of
    height attractors. This leads to a topological partition of the
    bifurcation diagram that gives the symbolic dynamic of all bifurcation
    diagram attractors with a unique template.
\end{abstract}

\noindent{\it Keywords}: R\"ossler system, bifurcation diagram, template, subtemplate

\section{Introduction}

Since 1976, the R\"ossler system \cite{rossler1976equation} is well know for
it's simplicity (three differential equations with only one non linear term)
and its dynamical richness producing chaos. Used as a basic system to
demonstrate various properties of dynamical systems, this system is still a
source inspiration for researchers. This system has been widely explored with
several tools. The main goal of this paper is to extend the use of the
topological characterization method to several chaotic attractors. We
introduce a way to use the template as a global description that contains
various attractors templates of the R\"ossler system.

In this paper, we will study this system in a parameter space to highlight
common properties of neighbours attractors in this space. Castro {\it et al.}
\cite{castro2007characterization} study of the parameter space of this system
using Lyapunov exponents reflects its dynamics (stable, chaotic or trajectory
diverging). The maps are built varying parameters $a$ and $c$ of the system.
These maps display fractal structure and illustrate period doubling cascades.
This principle is also employed by Barrio {\it et al.}
\cite{barrio2009qualitative} for the three parameters of the R\"ossler system.
Their analysis of local and global bifurcations of limit cycles permits to a
have a better understanding of the parameter space.  Additionally, attractors
with equilibrium points associated to their first return maps are also plotted
to illustrate the various dynamics of this system, including coexistence of
attractors. To enlarge this overview of recent work on the bifurcations and
dynamics on the R\"ossler system, Genesio {\it et al.}
\cite{genesio2008global} use the first-order harmonic balance technique to
study fold, flip and Neimark-Sacker bifurcations in the whole parameter space.
Finally, the recent work of Sprott \& Li \cite{sprott2015asymmetric} introduce
another way to reach coexisting attractors in addition to the cases identified
by Barrio {\it et al.} \cite{barrio2009qualitative}.

In this paper we study a bifurcation diagram of the R\"ossler system
exhibiting various dynamics using topological properties of attractors. We use
the topological characterization method based on the topological properties of
the attractor's periodic orbits \cite{gilmore2002topology}. The purpose is not
only to obtain templates of chaotic attractors but also to find common points
or properties as it as already been shown for this system by Letellier {\it et
al.} \cite{letellier1995unstable} for a ``funnel attractor''. In this
particular case, a linking matrix describes the template depending on the
number of branches. In this paper, we will explore a bifurcation diagram and
show that only one template contains all the templates of attractors as
subtemplates.

This paper is organized as follow. The first part introduces the Sprott \& Li
\cite{sprott2015asymmetric} work with their bifurcation diagram.  The second
part details the topological characterization method; height attractors are
studied and their templates are obtained. Then we prove that the eight
templates are subtemplates of a unique template. It describes the topological
structure of all the attractors of the entire bifurcation diagram.  Finally we
provide a partition of the bifurcation diagram giving the symbolic dynamic
associated with the unique template depending on the bifurcation parameter.

\section{Bifurcation diagram}

\begin{figure}[hbtp]
  \centering
  \includegraphics[width=.70\textwidth]{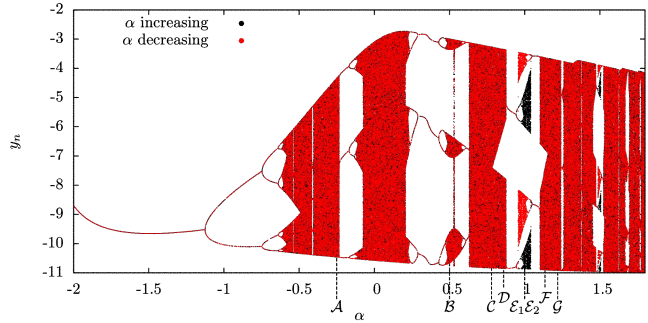}
  \vspace{-.5em}
  \caption{
    Bifurcation diagram when $\alpha$ varies: $\alpha$ increasing (red) and
    $\alpha$ decreasing (black). This figure reproduces the FIG.~4 of
    \cite{sprott2015asymmetric}. Using $\alpha$ increasing or
    decreasing replaces different initial conditions used in the original paper
    \cite{sprott2015asymmetric}. $\mathcal{A}$, $\mathcal{B}$, $\mathcal{C}$,
    $\mathcal{D}$, $\mathcal{E}_1$, $\mathcal{E}_2$, $\mathcal{F}$ and
    $\mathcal{G}$ refers to attractors solutions with the parameters indicated
    in \eqref{eq:alpha_value}.
  }
  \label{fig:rossler_bifurcation}
\end{figure}

Barrio \textit{et al.} \cite{barrio2009qualitative} highlight the fact that a
R\"ossler system can have to coexisting attractors as solutions for a set of
parameters.  Sprott \& Li \cite{sprott2015asymmetric} parametrize the R\"ossler
system \cite{rossler1976equation} with the parameter $\alpha$
\begin{equation}
  \left\{\begin{aligned}
    \dot{x} &= -y -z \\
    \dot{y} &= x +ay \\
    \dot{z} &= b+z(x-c)
  \end{aligned}\right.
  \text{ with }
  \left\{\begin{aligned}
    a &= 0.2 + 0.09\alpha \\
    b &= 0.2-0.06\alpha \\
    c &= 5.7-1.18\alpha
  \end{aligned}\right.
  \label{eq:rossler_alpha}
\end{equation}
in order to explore bifurcations. When $\alpha = 1$, two attractors coexist
in the phase space. We reproduce their bifurcation diagram when $\alpha$
varies. The value of the fixed
points of the system are
\begin{equation}
  S_\pm =
  \left|
    \begin{array}{l}
      \displaystyle x_\pm = \frac{c\pm \sqrt{c^2-4ab}}{2} \\[0.3cm]
      \displaystyle y_\pm = \frac{-c\mp \sqrt{c^2-4ab}}{2a} \\[0.3cm]
      \displaystyle z_\pm =  \frac{c \pm \sqrt{c^2-4ab}}{2a} \, .
    \end{array}
  \right.
  \label{eq:rossler_fixed_point}
\end{equation}
The bifurcation diagram Fig.~\ref{fig:rossler_bifurcation} is obtained using the following
Poincar\'e section
\begin{equation}
  \mathcal{P} \equiv \left\{ (y_n,-z_n) \in \mathbb{R}^2\ |
  -x_n = -x_- \right\}
  \label{eq:rossler_X_section}
\end{equation}
where $x_-$ is the $x$ value of the fixed point $S_-$ (see
\cite{rosalie2014toward} for details on this Poincar\'e section). The uses of
this Poincar\'e section explains why Fig.~\ref{fig:rossler_bifurcation} is
similar to FIG. 4 of \cite{sprott2015asymmetric}: we use $y_n$ and they
use $M$ which is a local maximum of $x$. Consequently, in both case, values
close to zero correspond to value close to the center of the attractor and
oppositely, high absolute values correspond to the outside boundary of the
attractor.

This diagram indicates parameter for the R\"ossler system where the solution
is a limit cycle or chaotic. This diagram exhibits a doubling period cascade
that is a classical route to chaos for $-2<\alpha<0.2$. This is followed by a
chaotic puff ($\alpha = 0.5$) and by various regimes (banded chaos and
almost fully developed chaos). We chose representative values of $\alpha$
where one ore two attractors are solutions of the system
\begin{equation}
  \begin{array}{ll|ll}
     \mathcal{A} & \alpha=-0.25& \mathcal{E}_1 & \alpha=1 \\
     \mathcal{B} & \alpha=0.5  & \mathcal{E}_2 & \alpha=1 \\
     \mathcal{C} & \alpha=0.78 & \mathcal{F} & \alpha=1.135 \\
     \mathcal{D} & \alpha=0.86 & \mathcal{G} & \alpha=1.22 \;.
  \end{array}
  \label{eq:alpha_value}
\end{equation}
We analyse these attractors using topological characterization method in order to
obtain a generic description of the attractors while $\alpha$ is varied.

\section{Topological characterization}

\begin{figure}[hbtp]
  \centering
  \begin{tabular}{cccccc}
    \multicolumn{2}{c}{Convention} &
    \multicolumn{2}{c}{Permutations} &
    \multicolumn{2}{c}{Torsions} \\
    \includegraphics[height=4em]{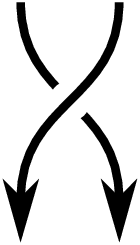} &
    \includegraphics[height=4em]{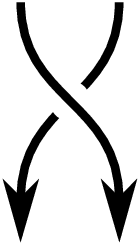} &
    \includegraphics[height=4em]{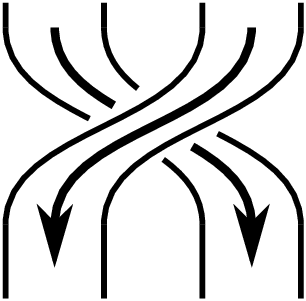} &
    \includegraphics[height=4em]{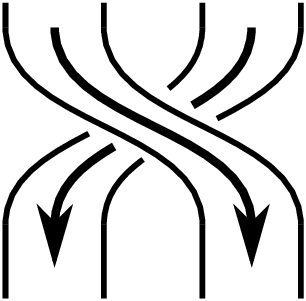} &
    \includegraphics[height=4em]{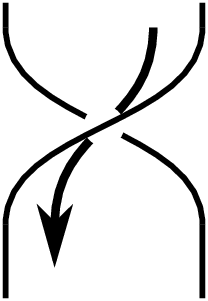} &
    \includegraphics[height=4em]{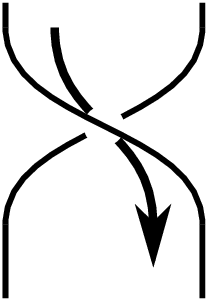} \\
    $+1$ &
    $-1$ &
    positive &
    negative &
    positive &
    negative
  \end{tabular}
  \vspace{-.5em}
  \caption{
    Convention representation of oriented crossings. The permutation
    between two branches is positive if the crossing generated is equal to
    $+1$, otherwise it is a negative permutation. We use the same
    convention for torsions.
  }
  \label{fig:convention}
\end{figure}

The main purpose of the topological characterization method is to build a
template using topological properties of periodic orbits. The template has
been introduced by \cite{birman1983knotted,franks1985entropy} further to the
works of Poincar\'e \cite{poincare1899methode}.  According to Ghrist {\it et
al.} \cite{ghrist1997knots}, a template is a compact branched two-manifold
with boundary and smooth expansive semiflow built locally from two type of
charts: joining and splitting. Each charts carries a semiflow, endowing the
template with an expanding semiflow, and the gluing maps between charts must
respect the semiflow and act linearly on the edges. This topological
characterization method is detailed by Gilmore \& Lefranc
\cite{gilmore1998topological,gilmore2002topology}. Recently, we detail
additional conventions to obtain templates that can be compared and sorted
\cite{rosalie2013systematic,rosalie2015systematic}. We start with a
brief description of the method. As the trajectories are chaotic, they are
unpredictable in a long term behavior. But attractors have a time invariant
global structure where its orbits compose its skeleton. The purpose of the
method is to use the topological properties of these orbits to describe the
structure of the attractor. We provide a sum up of this method with eight
steps (including our conventions):
\begin{enumerate}
  \item Display the attractor with a clockwise flow;
  \item Find the bounding torus;
  \item Build a Poincar\'e section;
\end{enumerate}
The first step permits to ensure that the study will be carried out to the
respect of conventions: clockwise flow having a clockwise toroidal boundary
and described by a clockwise template. This clockwise convention ensures us to
describe template with a unique linking matrix, a keystone to work only with
linking matrices \cite{rosalie2013systematic}. The toroidal boundary give a
global structure that permit to classify attractors.  For a given toroidal
boundary, a typical Poincar\'e section is associated according to the Tsankov
\& Gilmore theory \cite{tsankov2003strange}. This Poincar\'e section contains
one or more components to permit an effective discretization of trajectories
and consequently an efficient partition of the attractor.
\begin{enumerate}\setcounter{enumi}{3}
  \item Compute the first return map and define a symbolic dynamic;
  \item Extract and encode periodic orbits;
  \item Compute numerically the linking numbers between couple of orbits;
\end{enumerate}
The first return map details how two consecutive crossings of a trajectory
through the Poincar\'e section are related and permits to associate a symbol
to each point. It permits to define a partition of the attractor and a
symbolic dynamic. Associated symbols depend on the parity of the slope
(even for the increasing one, odd for the other). Up to this point, periodic
orbits structuring the attractor are extracted and encoded using this
symbolic dynamic. The linking number between a pair of orbits is a topological
invariant indicating how orbits are wind one around another. In this paper, we
use the orientation convention of Fig.~\ref{fig:convention}.

The final steps concern the template:
\begin{enumerate}\setcounter{enumi}{6}
  \item Propose a template;
  \item Validate the template with the theoretical computation of linking numbers.
\end{enumerate}
The template is clockwise oriented. The template of an attractor bounded by
genus one torus is defined by a unique linking matrix. This matrix describes how
branches are torn and permuted. We use the Melvin \& Tufillaro
\cite{melvin1994template} standard insertion convention: when the branches
stretch and squeeze, the left to the right order of the branches corresponds
to the bottom to top order. The diagonal elements of the matrix indicate the
torsions of branches and off diagonal elements give the permutations between
two branches. Finally, to validate a template, we use a procedure introduced
by Le Sceller {\it et al.} \cite{lesceller1994algebraic} that permits to compute linking
numbers theoretically from a linking matrix. Linking numbers obtained
theoretically with this method have to correspond with those obtained
numerically at the step (vi) to \emph{validate the template}. The challenge of
this procedure resides in the step (vii) because it is non trivial to find a
template whose theoretical linking numbers correspond to the numerically
computed linking numbers.

\subsection{Attractor $\mathcal{A}$}

\begin{figure}[h!]
  \centering
  \begin{tabular}{ccc}
    \includegraphics[width=.35\textwidth]{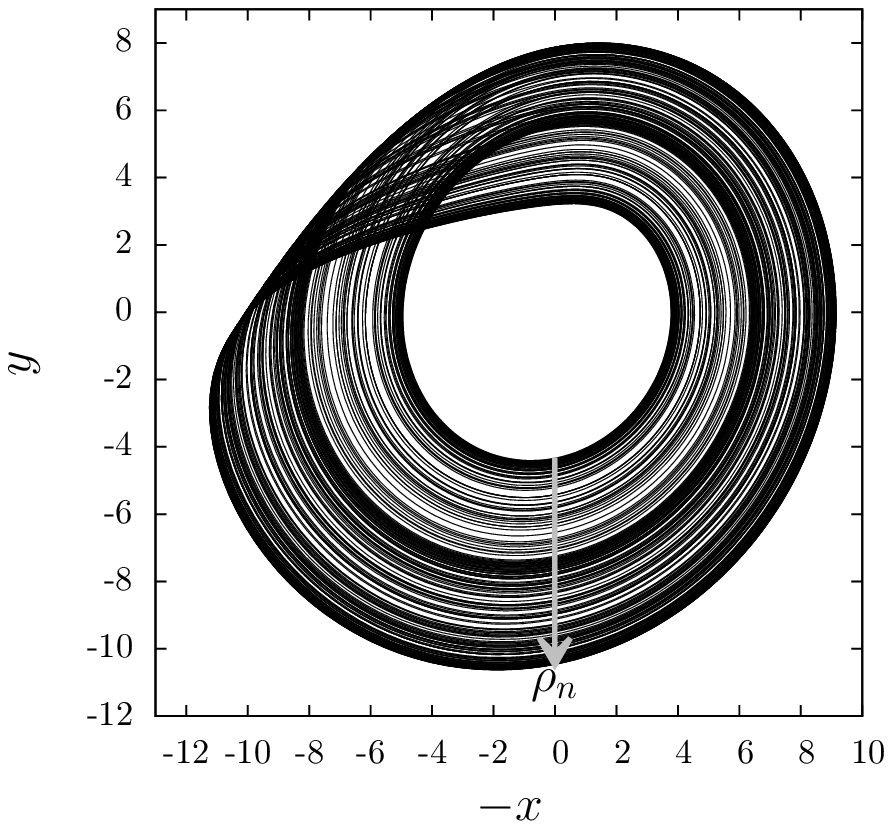} & &
    \includegraphics[width=.33\textwidth]{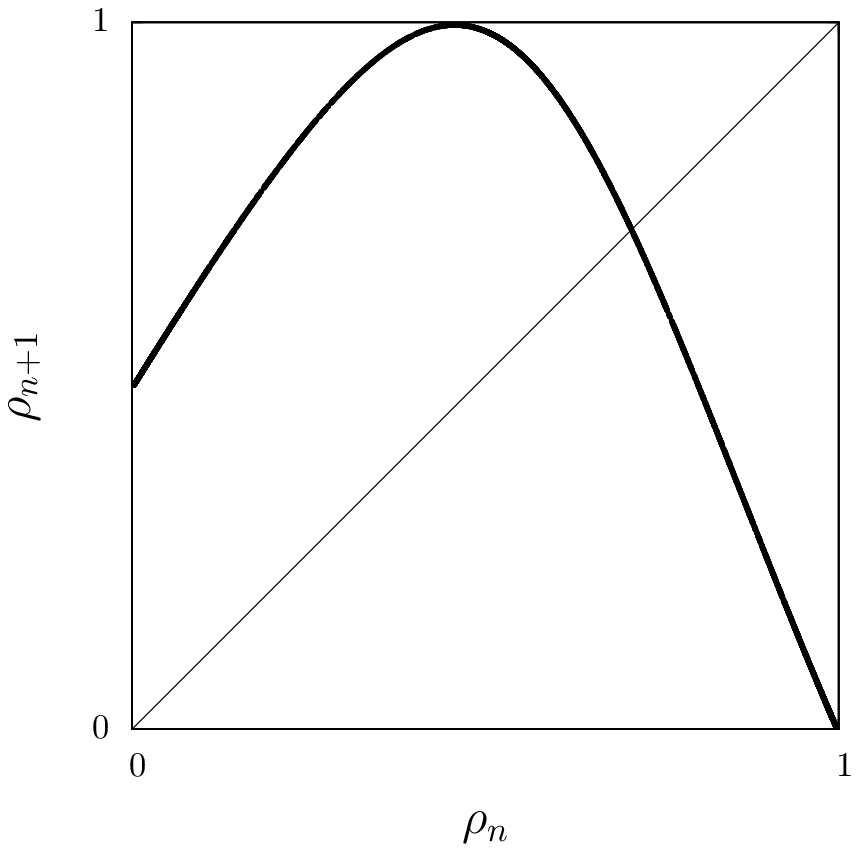}
    \put(-100,90){0}
    \put(-25,80){1}
    \\
    (a) Chaotic attractor $\mathcal{A}$ & &
    (b) First return map
  \end{tabular}
  \vspace{-.5em}
  \caption{
    (a) Chaotic attractor $\mathcal{A}$ solution to the equation
    \eqref{eq:rossler_alpha}. Parameter value $\alpha=-0.25$ with
    the initial conditions $x=-1.25 $, $y=-0.72 $ and $z=-0.1$. (b) First
    return map to the Poincar\'e section \eqref{eq:rossler_A_section} using
    $\rho_n$ (the arrow indicates the orientation).
  }
  \label{fig:rossler_A_attra_appli}
\end{figure}

In this section we will detail the previously described procedure
step by step for attractor $\mathcal{A}$.
\begin{enumerate}\setcounter{enumi}{0}
  \item Display the attractor with a clockwise flow;
\end{enumerate}
We propose to make a rotation of the attractor around the $y$-axis. Displaying
the attractor in the phase space $(-x,y)$, the flow evolves clockwise
(Fig.~\ref{fig:rossler_A_attra_appli}a).
\begin{enumerate}\setcounter{enumi}{1}
  \item Find the bounding torus;
\end{enumerate}
The attractor is bounded by a genus one torus: a surface with only one hole.
Consequently, a Poincar\'e section with one-component is required.
\begin{enumerate}\setcounter{enumi}{2}
  \item Build a Poincar\'e section;
\end{enumerate}
We build our Poincar\'e section using $x_-$ \eqref{eq:rossler_fixed_point}
\begin{equation}
  \mathcal{P} \equiv \left\{ (y_n,-z_n) \in \mathbb{R}^2\ |
  -x_n = -x_- \right\}\;.
  \label{eq:rossler_A_section}
\end{equation}
This Poincar\'e section
is a half-plan transversal to the flow as illustrated in grey
Fig.~\ref{fig:rossler_A_attra_appli}a.

\begin{figure}[ht]
  \centering
  \includegraphics[width=.5\textwidth]{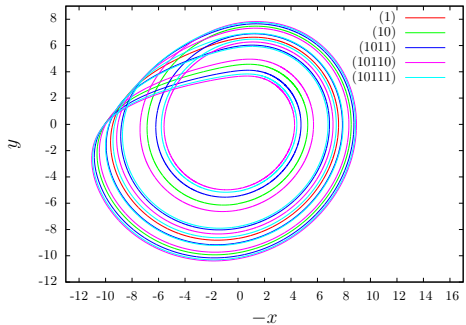}
  \vspace{-.5em}
  \caption{
    Orbits of the chaotic attractor $\mathcal{A}$.
  }
  \label{fig:rossler_A_orbits}
\end{figure}

\begin{enumerate}\setcounter{enumi}{3}
  \item Compute the first return map and define a symbolic dynamic;
\end{enumerate}
To compute the first return map, we first normalize the intersection value of
the flow through the Poincar\'e section: $\rho_n$. This value is oriented from
the inside to the outside (Fig.~\ref{fig:rossler_A_attra_appli}a). Then the
first return map is obtained by plotting $\rho_{n+1}$ versus $\rho_n$
(Fig.~\ref{fig:rossler_A_attra_appli}b). This return map is the classical
unimodal map made of an increasing branch followed by a decreasing one. This
first return map indicates that the classical ``horseshoe'' mechanism generate
this chaotic attractor. The symbolic dynamic is defined as follow: ``$0$'' for
the increasing branch and ``$1$'' for the decreasing one.
\begin{enumerate}\setcounter{enumi}{4}
  \item Extract and encode periodic orbits;
\end{enumerate}
Using, the first return map, we can locate periodic orbits that the flow
visits while it covers the attractor. For instance, there is only one period
one orbit because the bisector cross the map once. We extract five orbits with a period
lower than six: $1$, $10$, $1011$, $10110$ and $10111$
(Fig.~\ref{fig:rossler_A_orbits}).

\begin{enumerate}\setcounter{enumi}{5}
  \item Compute numerically the linking numbers;
\end{enumerate}
We compute linking numbers between each pair of periodic orbits. The linking
number between a pair of orbits is obtained numerically
(Tab.~\ref{tab:rossler_A_lk}).

\begin{table}[hb]
    \centering
    \caption{
        Linking numbers between pairs of orbits
        extracted from the symmetric chaotic attractor $\mathcal{A}$.
    }
    \resizebox{.4\textwidth}{!}{
        \begin{tabular}{ccccc}
            \\[-0.3cm]
            \hline \hline
            \\[-0.3cm]
            & (1) & (10) & (1011) & (10110)  \\[0.1cm]
            \hline
            \\[-0.3cm]
            (10)    & -1 & \\[0.1cm]
            (1011)  & -2 & -3 & \\[0.1cm]
            (10110) & -2 & -4 & -8  \\[0.1cm]
            (10111) & -2 & -4 & -8 & -10  \\[0.1cm]
            \hline \hline
        \end{tabular}
    }
    \vspace{-.5em}
    \label{tab:rossler_A_lk}
\end{table}

\begin{figure}[hbtp]
  \centering
  \includegraphics[height=.3\textheight]{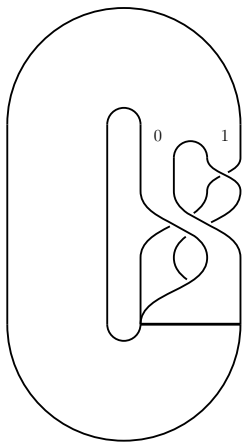}
  \vspace{-.5em}
  \caption{
      $T_\mathcal{A}$: template of attractor $\mathcal{A}$.
  }
  \label{fig:rossler_A_template}
\end{figure}

\begin{enumerate}\setcounter{enumi}{6}
  \item Propose a \emph{template};
\end{enumerate}
Using these linking numbers and the first return map structure, we propose the
template (Fig.~\ref{fig:rossler_A_template}). This template is made of trivial
part with a chaotic mechanism. The latter is composed by a splitting chart that
separate continuously the flow into two branches. The left one encoded ``$0$''
permute negatively over the right one encoded ``$1$''; the latter have a negative
torsion. After the torsion and permutation, branches stretch and squeeze to a
branch line using the standard insertion convention. This template is thus
described by the linking matrix
\begin{equation}
  T(\mathcal{A}) = \left[
    \begin{array}{C{1.3em}C{1.3em}}
       0 & -1  \\
      -1 & -1
    \end{array}
  \right\rsem\;.
\end{equation}

\begin{enumerate}\setcounter{enumi}{7}
  \item Validate the template computing theoretically the linking numbers.
\end{enumerate}
The theoretical calculus using the linking matrix and the orbits permits
to obtain the same table of linking numbers. This validates the template of
$\mathcal{A}$ defined by $T(\mathcal{A})$ (Fig.~\ref{fig:rossler_A_template}).

\subsection{Attractors $\mathcal{B}$ to $\mathcal{G}$}

\begin{figure}[htbp]
  \centering
  \begin{tabular}{ccc}
    \includegraphics[width=.3\textwidth]{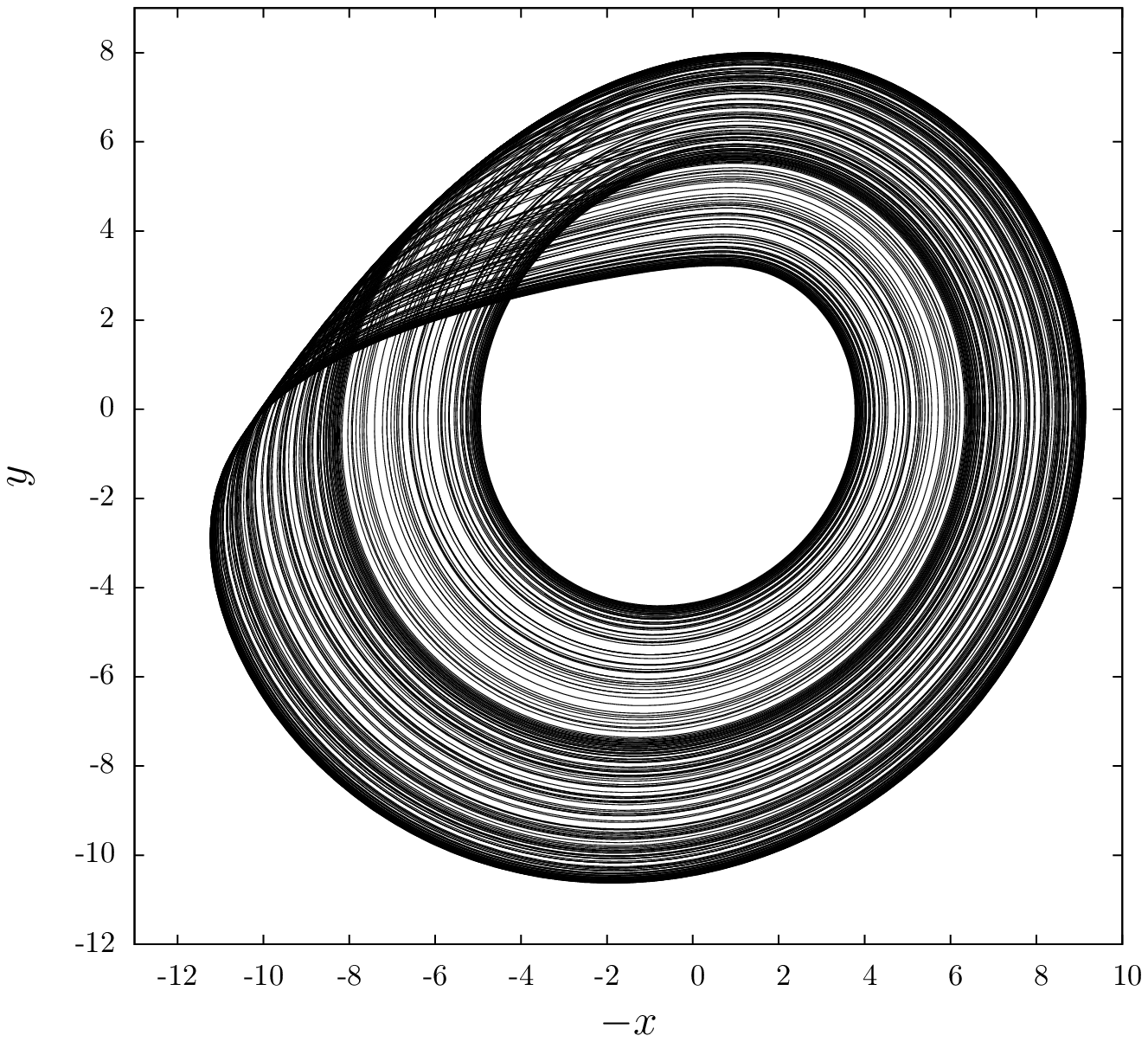} &
    \includegraphics[width=.3\textwidth]{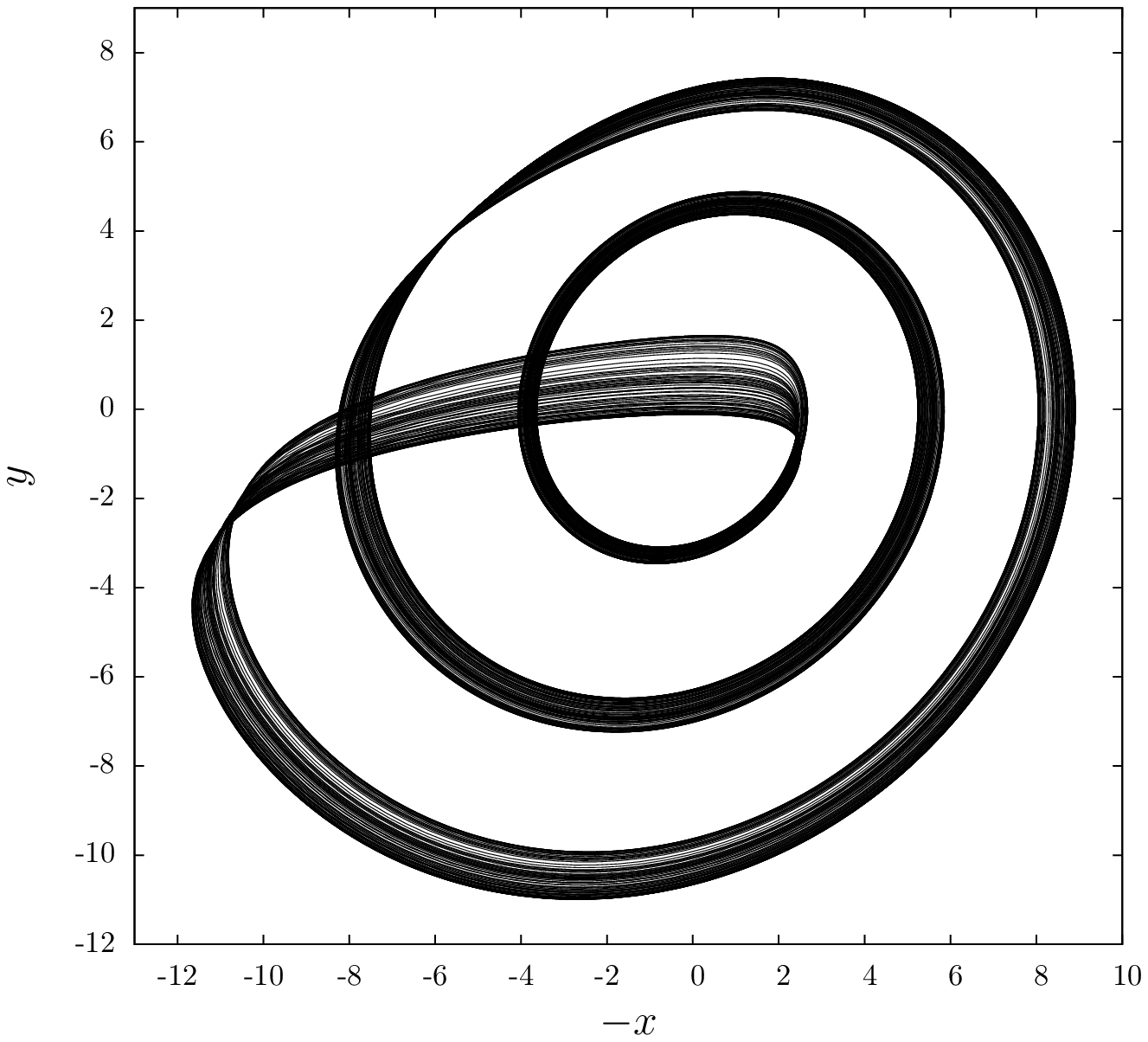} &
  \includegraphics[width=.3\textwidth]{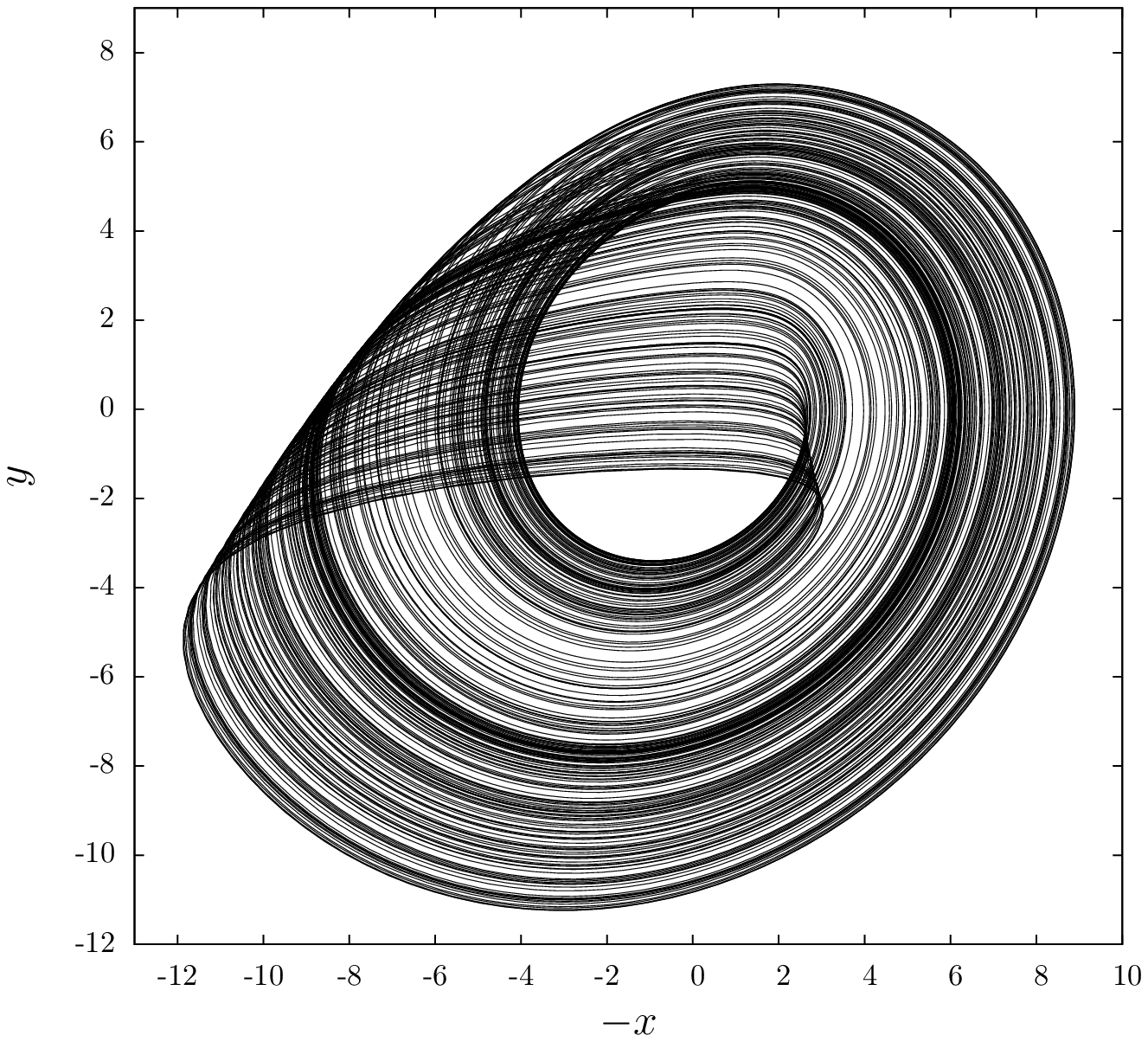} \\
  (a) $\mathcal{A}$ for $\alpha=-0.25$ &
  (b) $\mathcal{B}$ for $\alpha=0.5$ &
  (c) $\mathcal{C}$ for $\alpha=0.78$ \\[.5em]
  \includegraphics[width=.3\textwidth]{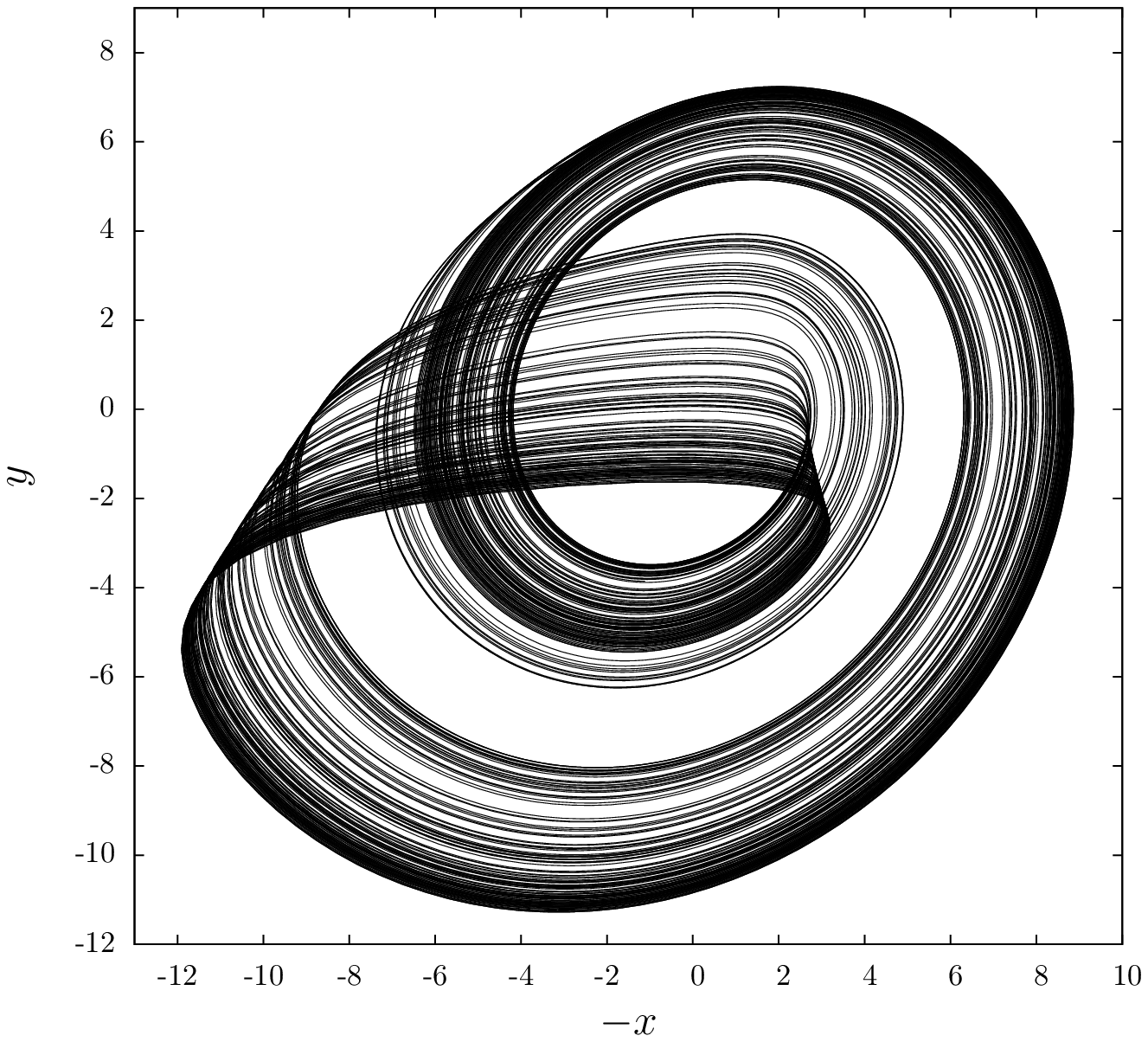} &
  \includegraphics[width=.3\textwidth]{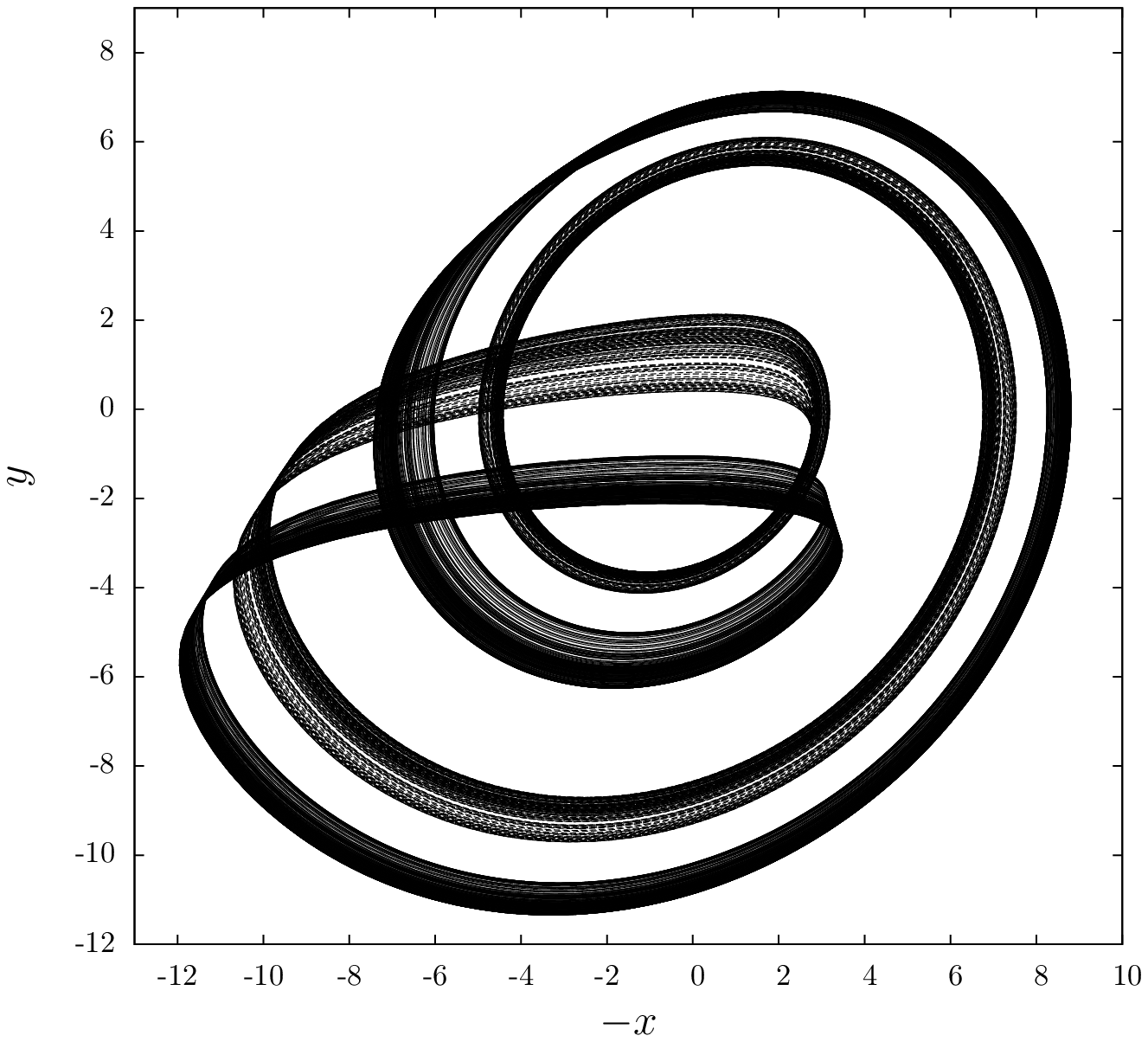} &
  \includegraphics[width=.3\textwidth]{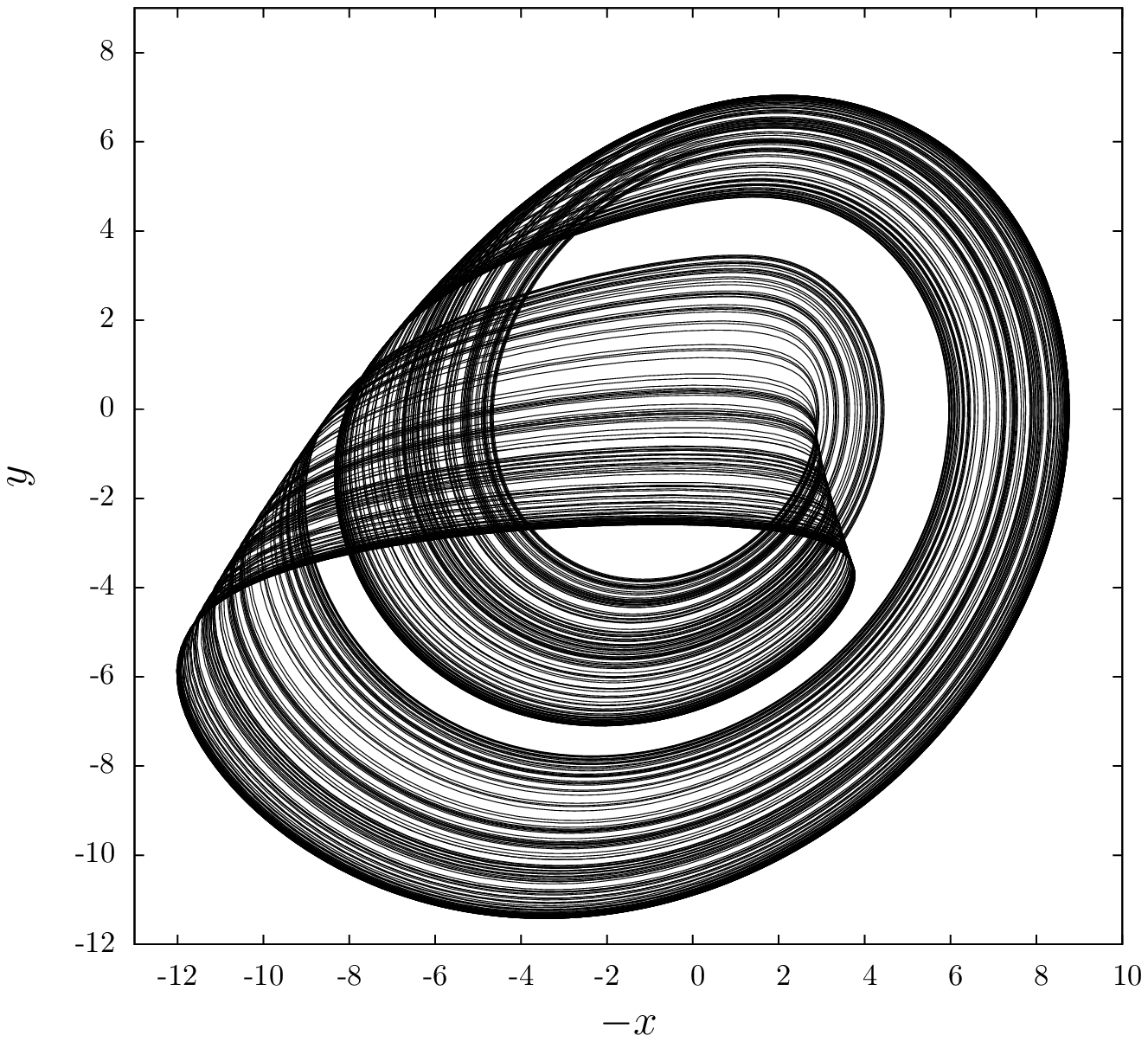} \\
  (d) $\mathcal{D}$ for $\alpha=0.86$ &
  (e) $\mathcal{E}_1$ and $\mathcal{E}_2$ for $\alpha=1$ &
  (f) $\mathcal{F}$ for $\alpha=1.135$ \\[.5em]
  & \includegraphics[width=.3\textwidth]{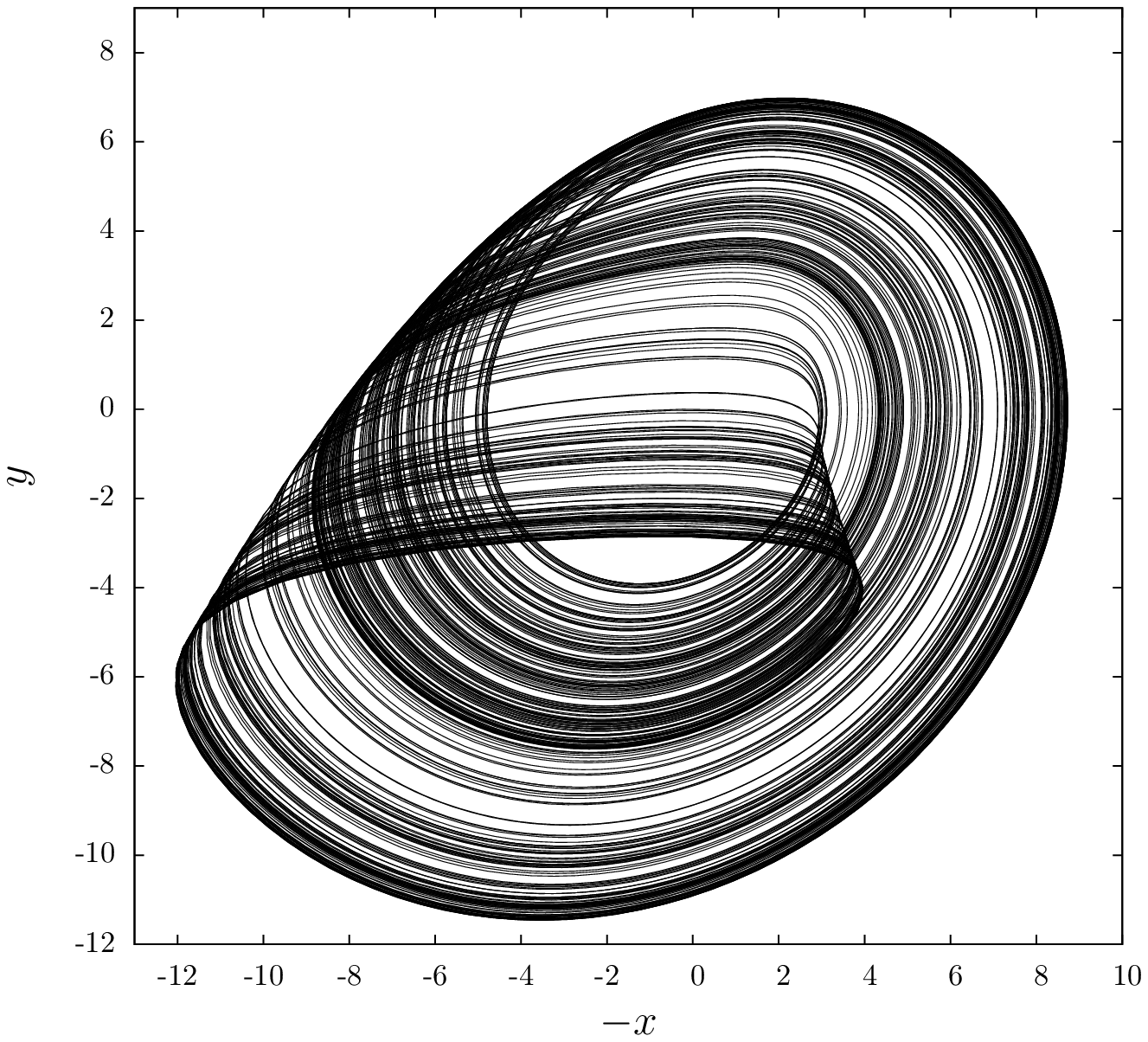} & \\
  & (g) $\mathcal{G}$ for $\alpha=1.22$ &
  \end{tabular}
  \vspace{-.5em}
  \caption{
    Eight chaotic attractors for different values of $\alpha$ from the
    bifurcation diagram (Fig.~\ref{fig:rossler_bifurcation})
  }
  \label{fig:rossler_A_to_G_attra}
\end{figure}

In this section, we will only give the key steps for others attractors:
$\mathcal{B}$, $\mathcal{C}$, $\mathcal{D}$, $\mathcal{E}_1$, $\mathcal{E}_2$,
$\mathcal{F}$ and $\mathcal{G}$. We start with the
Fig.~\ref{fig:rossler_A_to_G_attra} displaying these attractors for parameters
\eqref{eq:alpha_value} and with a clockwise flow evolution. We can observe
that these attractors are made of fully developed
chaos (Fig.~\ref{fig:rossler_A_to_G_attra}acg) or banded chaos
(Fig.~\ref{fig:rossler_A_to_G_attra}bdf) or coexisting attractors of banded
chaos (Fig.~\ref{fig:rossler_A_to_G_attra}e).

\begin{figure}[htbp]
  \centering
  \begin{tabular}{ccc}
    \includegraphics[width=.3\textwidth]{rossler_A_appli.eps}
    \put(-90,80){0}
    \put(-20,70){1}
    &
    \includegraphics[width=.3\textwidth]{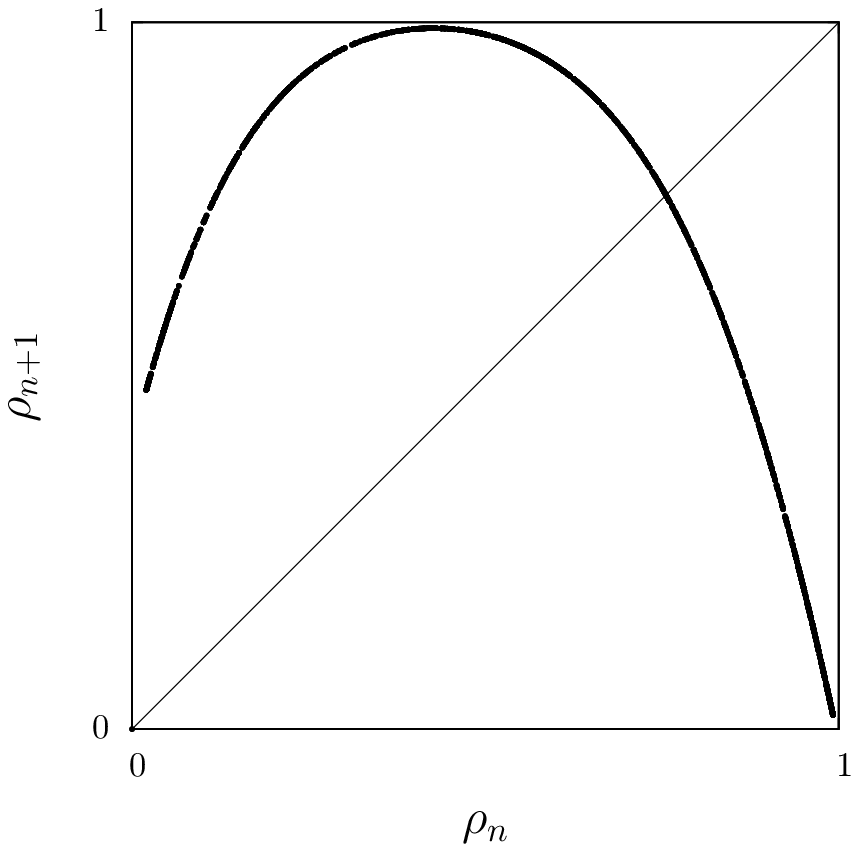}
    \put(-90,90){0}
    \put(-15,70){1}
     &
  \includegraphics[width=.3\textwidth]{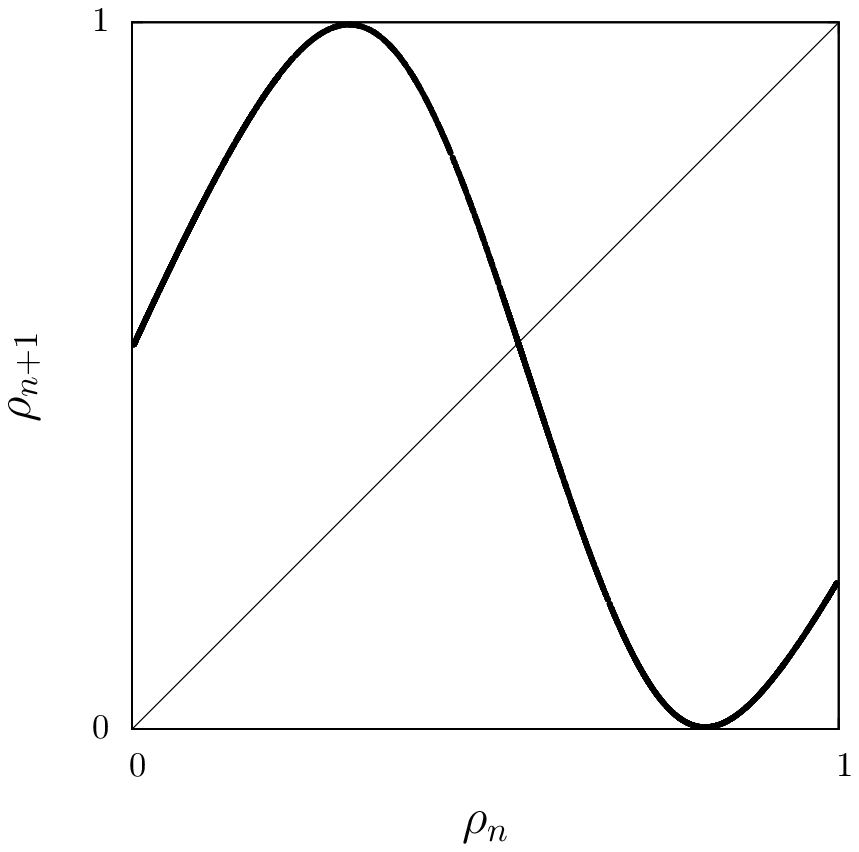}
    \put(-90,90){0}
    \put(-40,60){1}
    \put(-15,30){2}
     \\
  (a) $\mathcal{A}$ for $\alpha=-0.25$ &
  (b) $\mathcal{B}$ for $\alpha=0.5$ &
  (c) $\mathcal{C}$ for $\alpha=0.78$ \\
  \includegraphics[width=.3\textwidth]{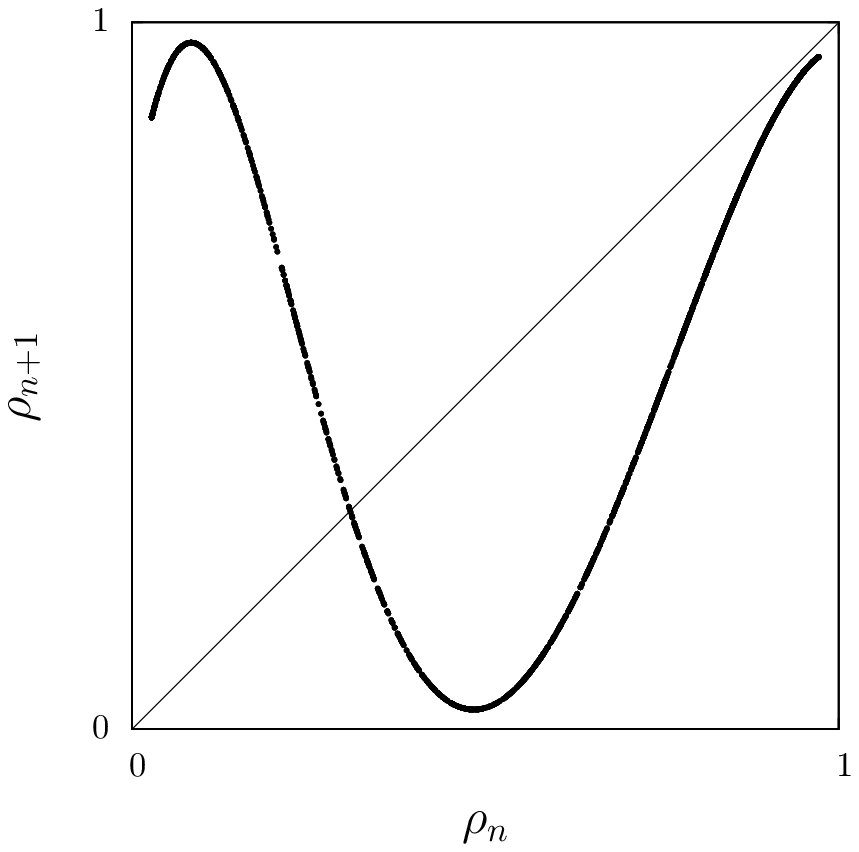}
    \put(-90,90){0}
    \put(-70,70){1}
    \put(-15,70){2}
    &
  \includegraphics[width=.3\textwidth]{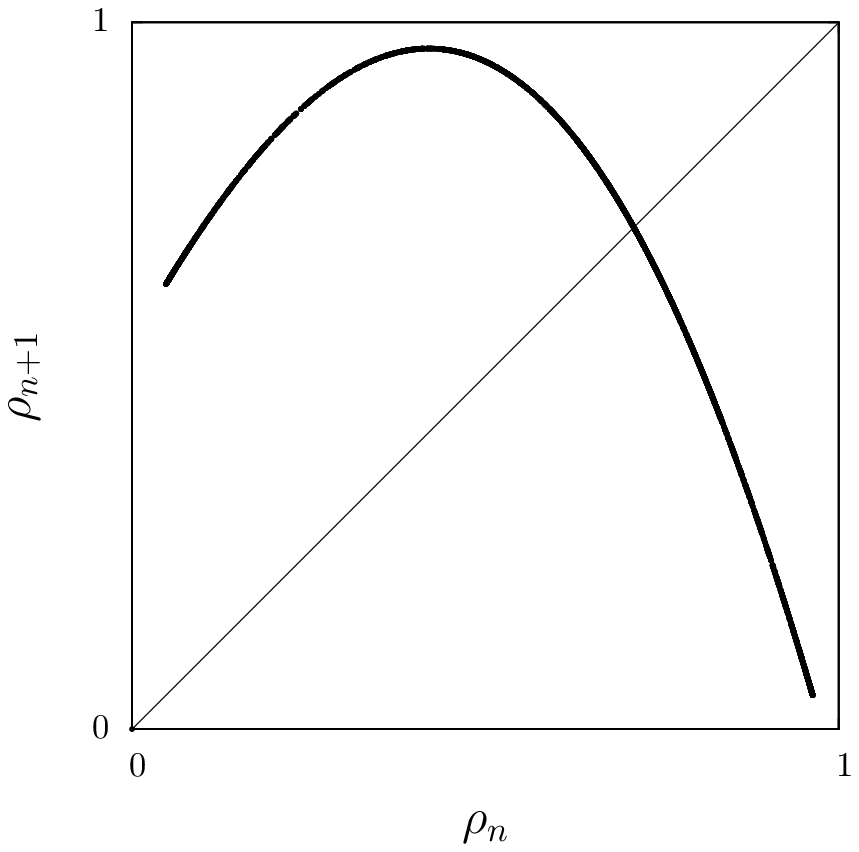}
    \put(-90,90){0}
    \put(-20,70){1}
     &
  \includegraphics[width=.3\textwidth]{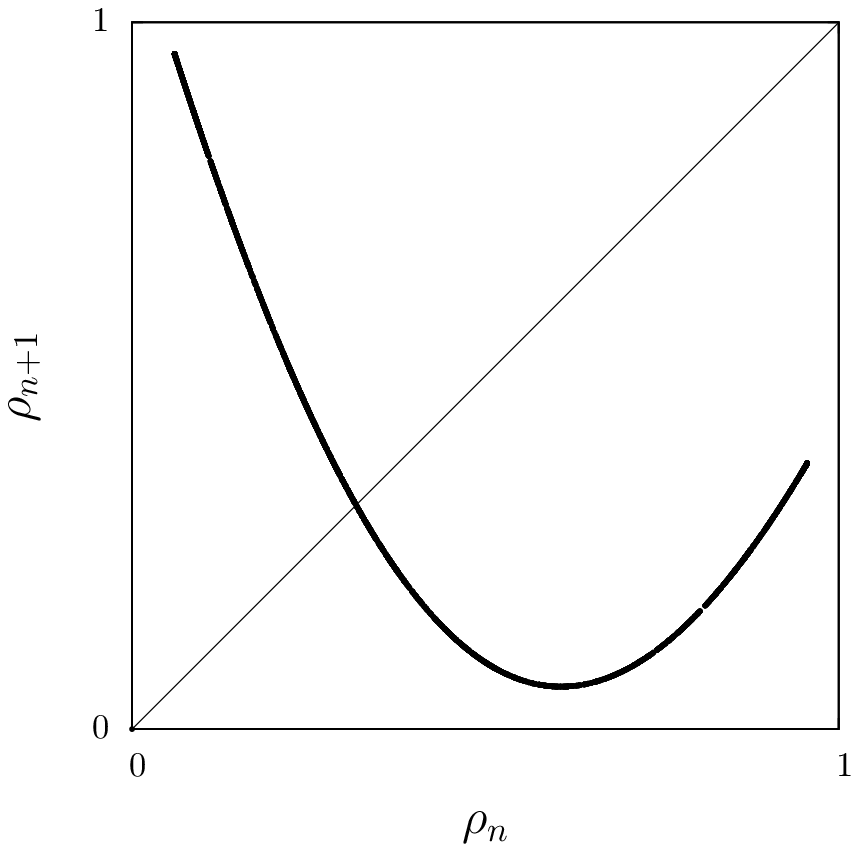}
    \put(-70,70){1}
    \put(-20,40){2}
     \\
  (d)    $\mathcal{D}$ for  $\alpha=0.86$ &
  (e$_1$) $\mathcal{E}_1$ for $\alpha=1$ &
  (e$_2$) $\mathcal{E}_2$ for $\alpha=1$ \\
  \includegraphics[width=.3\textwidth]{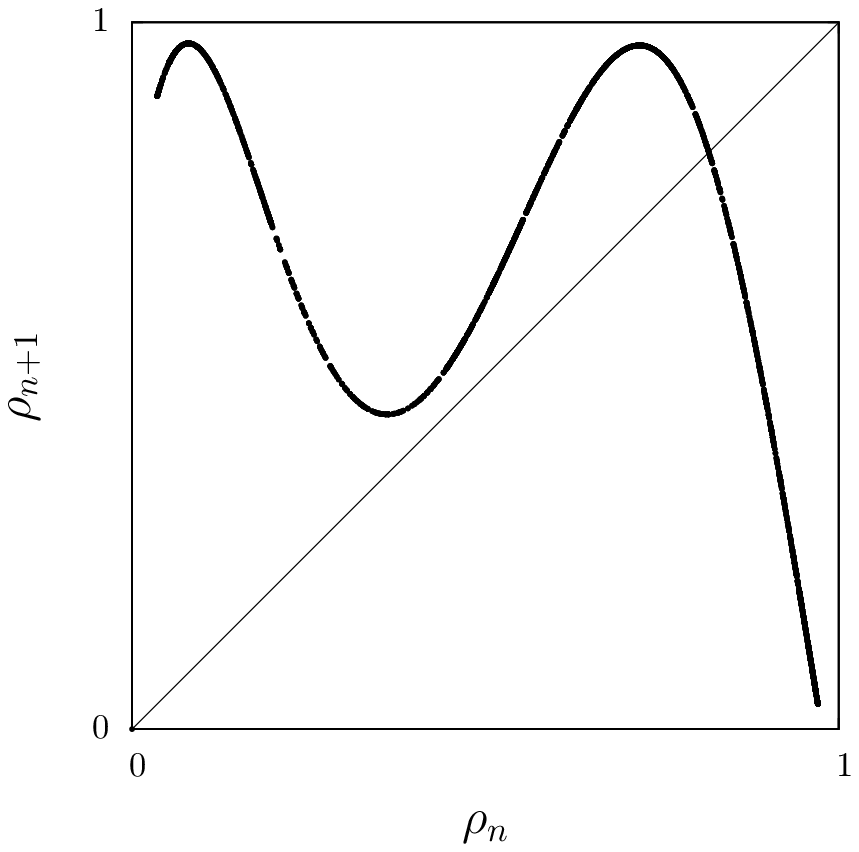}
    \put(-90,90){0}
    \put(-73,80){1}
    \put(-53,80){2}
    \put(-23,70){3}
     &
   \includegraphics[width=.3\textwidth]{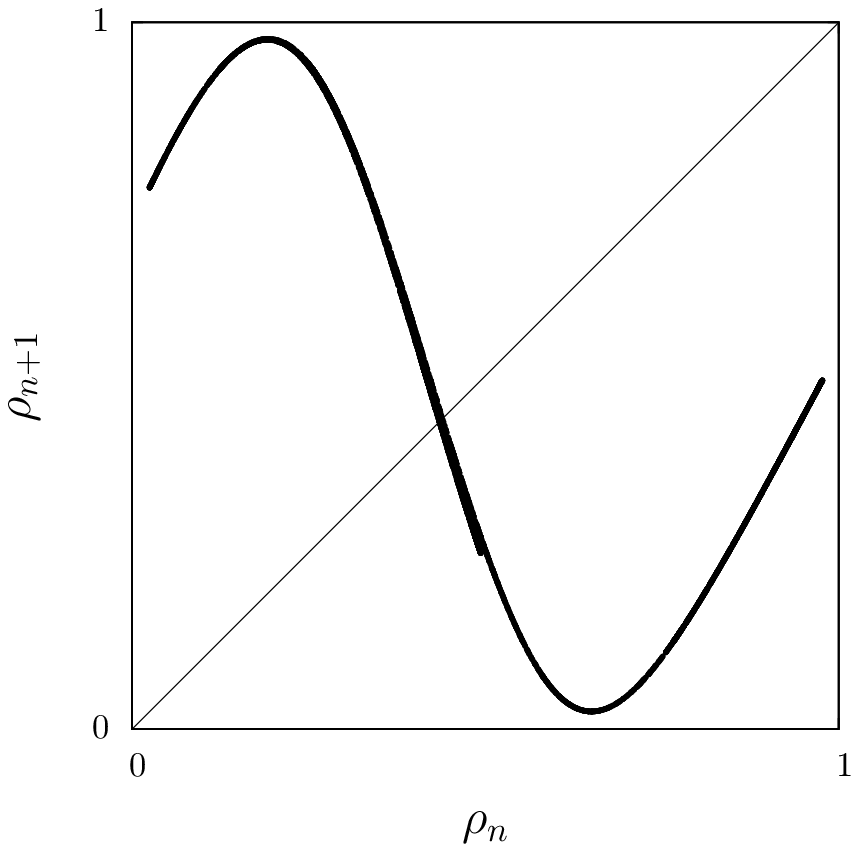}
    \put(-90,80){0}
    \put(-50,70){1}
    \put(-20,45){2}
     & \\
  (f) $\mathcal{F}$ for $\alpha=1.135$ &
  (g) $\mathcal{G}$ for $\alpha=1.22$ &
  \end{tabular}
  \vspace{-.5em}
  \caption{
    First return maps of the eight attractors of the
    Fig.~\ref{fig:rossler_A_to_G_attra}.
  }
  \label{fig:rossler_A_to_G_appli}
\end{figure}

All these attractors are bounded by a genus one torus. We use the
Poincar\'e sections: \eqref{eq:rossler_A_C_G_section} for $\mathcal{A}$,
$\mathcal{C}$, $\mathcal{G}$, \eqref{eq:rossler_D_E_F_section}
$\mathcal{D}$, $\mathcal{E}_1$, $\mathcal{E}_2$, $\mathcal{F}$ and
\eqref{eq:rossler_B_section} for $\mathcal{B}$:
\begin{equation}
  \mathcal{P} \equiv \left\{ (y_n,-z_n) \in \mathbb{R}^2\ |
  -x_n = -x_- \right\}\ ,
  \label{eq:rossler_A_C_G_section}
\end{equation}
\begin{equation}
  \mathcal{P} \equiv \left\{ (y_n,-z_n) \in \mathbb{R}^2\ |
  -x_n = -x_- , \ -\dot{x}_n<0,y<-7\right\}\ ,
  \label{eq:rossler_D_E_F_section}
\end{equation}
\begin{equation}
  \mathcal{P} \equiv \left\{ (y_n,-z_n) \in \mathbb{R}^2\ |
  -x_n = -x_- , \ -\dot{x}_n<0,y<-9\right\}\;.
  \label{eq:rossler_B_section}
\end{equation}
We compute the first return maps to these Poincar\'e sections using a
normalized variable $\rho_n$ oriented from the inside to the outside of the
boundary. The eight return maps (Fig.~\ref{fig:rossler_A_to_G_appli})
are multimodal with differential points and the number
of their branches are two, three or four. We chose a symbolic dynamic for each
first return map with respect to the slope orientation of branches
\begin{equation}
  \begin{array}{ll|ll}
    \mathcal{A} & 0\ 1 & \mathcal{E}_1 & 0\ 1  \\
    \mathcal{B} & 0\ 1 & \mathcal{E}_2 & 1\ 2  \\
    \mathcal{C} & 0\ 1\ 2& \mathcal{F} & 0\ 1\ 2\ 3   \\
    \mathcal{D} & 0\ 1\ 2& \mathcal{G} & 0\ 1\ 2 \;.
  \end{array}
  \label{eq:symbolic_dynamic}
\end{equation}

We extract a set of orbits for each attractors and numerically compute linking
numbers between pairs of orbits (Tab.~\ref{tab:rossler_A-G_lk_template}).
Thus, we propose templates. They are validated using Le Sceller {\it et al.}
procedure \cite{lesceller1994algebraic}. All the results are summed up in the
Tab.~\ref{tab:rossler_A-G_lk_template}.

\begin{table}[p]
  \centering
  \caption{
    Linking numbers between pairs of orbits extracted from the attractors of
    Fig.~\ref{fig:rossler_A_to_G_attra} and their associated linking matrix
    describing their template.
  }
  \vspace{.5em}
\begin{tabular}{clc}
    & \hfill Linking numbers \hfill\null& Linking matrix\\[-1em]\\
  $\mathcal{A}$ &
  {\footnotesize
    \begin{tabular}{ccccc}
      \hline
      & (1) & (10) & (1011) & (10110)  \\[0.1cm]
      \hline
      \\[-0.3cm]
      (10)    & -1 & \\[0.1cm]
      (1011)  & -2 & -3 & \\[0.1cm]
      (10110) & -2 & -4 & -8  \\[0.1cm]
      (10111) & -2 & -4 & -8 & -10  \\[0.1cm]
      \hline 
  \end{tabular}}
  &
  $T(\mathcal{A}) = \left[
    \begin{array}{C{1.3em}C{1.3em}}
       0 & -1  \\
      -1 & -1
    \end{array}
    \right\rsem$ \\[1cm]
  $\mathcal{B}$ &
  {\footnotesize
  \begin{tabular}{ccccc}
    \hline 
    & (1) & (10) & (1011) & (10110)  \\[0.1cm]
    \hline
    \\[-0.3cm]
    (10)    & -5 & \\[0.1cm]
    (1011)  & -10 & -21 & \\[0.1cm]
    (10110) & -13 & -26  & -52   \\[0.1cm]
    (10111) & -13 & -26  & -52 & -65  \\[0.1cm]
    \hline 
\end{tabular}}&
  $T(\mathcal{B}) = \left[
    \begin{array}{C{1.3em}C{1.3em}}
      -6 & -6  \\
      -6 & -5
    \end{array}
  \right\rsem$ \\[1cm]
  $\mathcal{C}$ &
  {\footnotesize
  \begin{tabular}{ccccc}
    \hline
       & (1) & (10) & (2010) & (2011)  \\[0.1cm]
       \hline
       \\[-0.3cm]
       (10)   & -1 & \\[0.1cm]
       (2010)  & -2 & -3 & \\[0.1cm]
       (2011)  & -2 & -3 & -6  \\[0.1cm]
       (1011) & -2 & -3 & -6 & -6  \\[0.1cm]
       \hline 
     \end{tabular}
  }&
  $
T(\mathcal{C}) = \left[
    \begin{array}{C{1.3em}C{1.3em}C{1.3em}}
       0 & -1 & -1  \\
      -1 & -1 & -1  \\
      -1 & -1 &  0
    \end{array}
  \right\rsem
  $ \\[1cm]
  $\mathcal{D}$ &
  {\footnotesize
  \begin{tabular}{cccccc}
    \hline
       & (1) & (221) & (211) & (2221) & (2211)  \\[0.1cm]
       \hline
       \\[-0.3cm]
       (221)   & -4 & \\[0.1cm]
       (211)   & -4 & -12 & \\[0.1cm]
       (2221)  & -5 & -15 & -15  \\[0.1cm]
       (2211)  & -5 & -15 & -15 & -20  \\[0.1cm]
       (22110) & -6 & -18 & -18 & -24 &-24 \\[0.1cm]
       \hline 
     \end{tabular}
  }&
  $
T(\mathcal{D}) = \left[
    \begin{array}{C{1.3em}C{1.3em}C{1.3em}}
      -4 & -4 & -4 \\
      -4 & -3 & -3 \\
      -4 & -3 & -2
    \end{array}
  \right\rsem
  $ \\[1.2cm]
  $\mathcal{E}_1$ &
  {\footnotesize
    \begin{tabular}{ccccc}
      \hline
       & (1) & (10) & (1011) & (101111)  \\[0.1cm]
       \hline
       \\[-0.3cm]
       (10)   & -3 & \\[0.1cm]
       (1011)  & -6 & -11 & \\[0.1cm]
       (101111)  & -9 & -17 & -34  \\[0.1cm]
       (101110) & -9 & -17 & -34 & -51  \\[0.1cm]
       \hline 
     \end{tabular}
  }&
  $
  T(\mathcal{E}_1) = \left[
    \begin{array}{C{1.3em}C{1.3em}}
      -2 & -3  \\
      -3 & -3
    \end{array}
  \right\rsem
  $ \\[1cm]
  $\mathcal{E}_2$ &
  {\footnotesize
    \begin{tabular}{ccccc}
      \hline
       & (1) & (21) & (2111) & (211111)  \\[0.1cm]
       \hline
       \\[-0.3cm]
       (21)   & -3 & \\[0.1cm]
       (2111)  & -6 & -11 & \\[0.1cm]
       (211111)  & -9 & -17 & -34  \\[0.1cm]
       (212111) & -9 & -17 & -34 & -51  \\[0.1cm]
       \hline 
     \end{tabular}
  }&
  $
  T(\mathcal{E}_2) = \left[
    \begin{array}{C{1.3em}C{1.3em}}
      -3 & -3  \\
      -3 & -2
    \end{array}
  \right\rsem
  $ \\[1cm]
  $\mathcal{F}$ &
  {\footnotesize
    \begin{tabular}{cccccccc}
      \hline
       & (3) & (30) & (31) & (32) & (313) & (312) & (322)  \\[0.1cm]
       \hline
       \\[-0.3cm]
       (30)  & -3 & \\[0.1cm]
       (31)  & -3 & -6 & \\[0.1cm]
       (32)  & -3 & -6 & -6  \\[0.1cm]
       (313) & -4 & -9 & -9 & -8   \\[0.1cm]
       (312) & -4 & -9 & -9 & -8 & -12  \\[0.1cm]
       (322) & -4 & -9 & -9 & -8 & -12 & -12  \\[0.1cm]
       (332) & -4 & -9 & -9 & -8 & -12 & -12 & -12  \\[0.1cm]
       \hline
     \end{tabular}
  }&
  $
  T(\mathcal{F}) = \left[
    \begin{array}{C{1.3em}C{1.3em}C{1.3em}C{1.3em}}
      -4 & -4 & -4 & -4  \\
      -4 & -3 & -3 & -3  \\
      -4 & -3 & -2 & -3  \\
      -4 & -3 & -3 & -3
    \end{array}
  \right\rsem
  $ \\[1.7cm]
  $\mathcal{G}$ &
  {\footnotesize
    \begin{tabular}{ccccc}
      \hline
       & (1) & (21) & (2111) & (2021)  \\[0.1cm]
       \hline
       \\[-0.3cm]
       (21)   & -1 & \\[0.1cm]
       (2111) & -2 & -3 & \\[0.1cm]
       (2021) & -2 & -3 & -6  \\[0.1cm]
       (2110) & -2 & -3 & -6 & -6  \\[0.1cm]
       \hline 
     \end{tabular}
  }&
  $
T(\mathcal{G}) = \left[
    \begin{array}{C{1.3em}C{1.3em}C{1.3em}}
       0 & -1 & -1  \\
      -1 & -1 & -1  \\
      -1 & -1 &  0
    \end{array}
  \right\rsem
  $
\end{tabular}
  \label{tab:rossler_A-G_lk_template}
\end{table}

Only attractors $\mathcal{C}$ and $\mathcal{G}$ have the same template even if
they have not the same orbits of period lower than five.
The dynamic is not fully developed on each branch: some orbits are
missing in both attractors compare to the full set of orbits. On the other
hand, the two coexisting attractors have the same linking numbers but the
orbits encode in two distinct ways. Their linking matrix are related as their
template too. In fact, the orientation of the chaotic mechanism is the
opposite one. This case permits to underline how orientation conventions are
necessary to distinguish these two attractors.

\section{Subtemplate}

\subsection{Algebraical relation between linking matrix}

In this section, we will use some algebraic relation between linking matrix
already defined in our previous papers
\cite{rosalie2013systematic,rosalie2015systematic}. Here, we provide an
overview of these relations. In the following description, a \emph{strip} also
nominates a branch of a branched manifold. First of all, a \emph{linker} is a
synthesis of the relative organization of $n$ strips: torsions and
permutations in a planar projection (Fig.~\ref{fig:convention}). A
\emph{mixer} is a linker ended by a joining chart that stretch and squeeze
strips to a branch line. In the previous section, templates are only
composed by one mixer defined by a linking matrix. We also
define the concatenation of a torsion with a mixer and the concatenation of
two mixers (see \cite{rosalie2013systematic,rosalie2015systematic} for more
details). We remind that $\mathcal{X}$ designates the attractor, $T_\mathcal{X}$
its template and $T(\mathcal{X})$ the linking matrix that define its template.
Using these algebraical relations between mixers and torsion, we can link the
mixers of previously studied attractors:
\begin{equation}
  \begin{array}{ll}
    T(\mathcal{C}) = T(\mathcal{G}) &  \\[0.5cm]
    T(\mathcal{E}_1) = T(\mathcal{E}_2)^p &
      \left[
    \begin{array}{C{1.3em}C{1.3em}}
      -3 & -3  \\
      -3 & -2
    \end{array}
    \right\rsem =
    \left[
      \begin{array}{C{1.3em}C{1.3em}}
        -2 & -3  \\
        -3 & -3
      \end{array}
      \right\rsem^p
    \\[0.5cm]

  T(\mathcal{B}) = [-5] \oplus T(\mathcal{A}) &
  \left[
    \begin{array}{C{1.3em}C{1.3em}}
      -6 & -6  \\
      -6 & -5
    \end{array}
    \right\rsem =
    \left[
      \begin{array}{C{1.3em}C{1.3em}}
        -5 & -5  \\
        -5 & -5
      \end{array}
    \right]
    +
    \left[
      \begin{array}{C{1.3em}C{1.3em}}
        0 & -1  \\
        -1 & -1
      \end{array}
      \right\rsem\\[0.5cm]
  T(\mathcal{E}_2) = [-2] \oplus T(\mathcal{A}) &
  \left[
    \begin{array}{C{1.3em}C{1.3em}}
      -2 & -3  \\
      -3 & -3
    \end{array}
    \right\rsem =
    \left[
      \begin{array}{C{1.3em}C{1.3em}}
        -2 & -2  \\
        -2 & -2
      \end{array}
    \right]
    +
    \left[
      \begin{array}{C{1.3em}C{1.3em}}
        0 & -1  \\
        -1 & -1
      \end{array}
      \right\rsem
  \end{array}
  \label{eq:torsion_mixer}
\end{equation}

\subsection{Subtemplates}

A subtemplate is defined as follow by Ghrist {\it et al.}
\cite{ghrist1997knots}: a \emph{subtemplate} $\mathcal{S}$ of a template
$\mathcal{T}$, written $\mathcal{S} \subset \mathcal{T}$, is a topological
subset of $\mathcal{T}$ which, equipped with the restriction of a semiflow of
$\mathcal{T}$ to $\mathcal{S}$, satisfies the definition of a template.
For the eight attractors previously studied ($\mathcal{A}$, $\mathcal{B}$,
$\mathcal{C}$, $\mathcal{D}$, $\mathcal{E}_1$, $\mathcal{E}_2$, $\mathcal{F}$
and $\mathcal{G}$) we will demonstrate that their templates are subtemplate of
the template of $\mathcal{C}$ made of one mixer defined by
\begin{equation}
\left[
    \begin{array}{C{1.3em}C{1.3em}C{1.3em}}
       0 & -1 & -1  \\
      -1 & -1 & -1  \\
      -1 & -1 &  0
    \end{array}
  \right\rsem\;.
\end{equation}

Using the linking matrix defining $T_\mathcal{A}$ and $T_\mathcal{C}$, we
directly find that $T(\mathcal{A})$ is a subset of $T(\mathcal{C})$ with the
two first lines and columns
\begin{equation}
    \left[
    \begin{array}{C{1.3em}C{1.3em}}
       0 & -1  \\
      -1 & -1
    \end{array}
    \right\rsem
    \subset
\left[
    \begin{array}{C{1.3em}C{1.3em}C{1.3em}}
       0 & -1 & -1  \\
      -1 & -1 & -1  \\
      -1 & -1 &  0
    \end{array}
  \right\rsem\;.
\end{equation}

\begin{figure}[hbtp]
  \centering
  \begin{tabular}{cc}
    \includegraphics[height=.3\textheight]{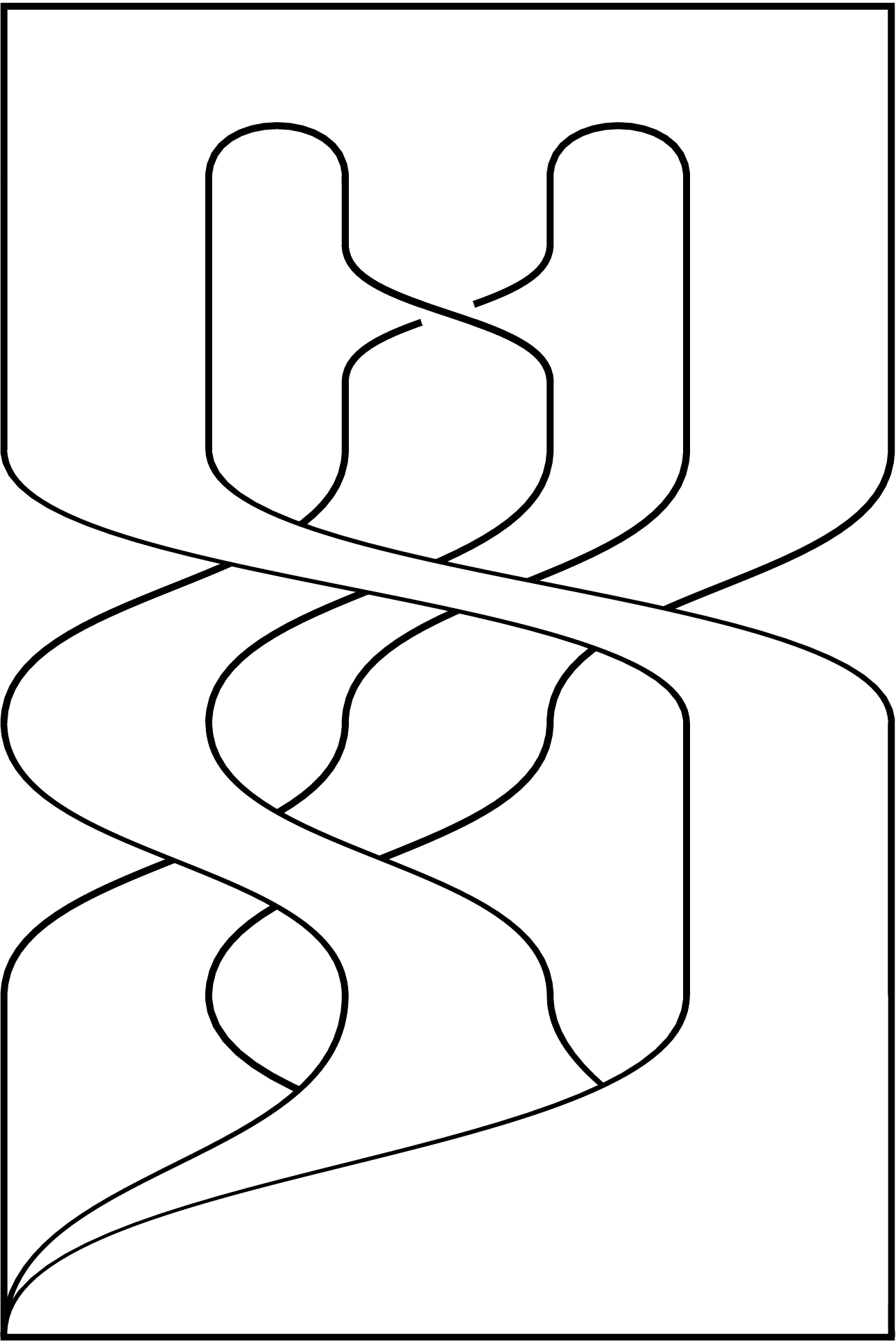}
    \put(-105,150){\large 0}
    \put(-63,150){\large 1}
    \put(-18,150){\large 2}
    &
    \includegraphics[height=.3\textheight]{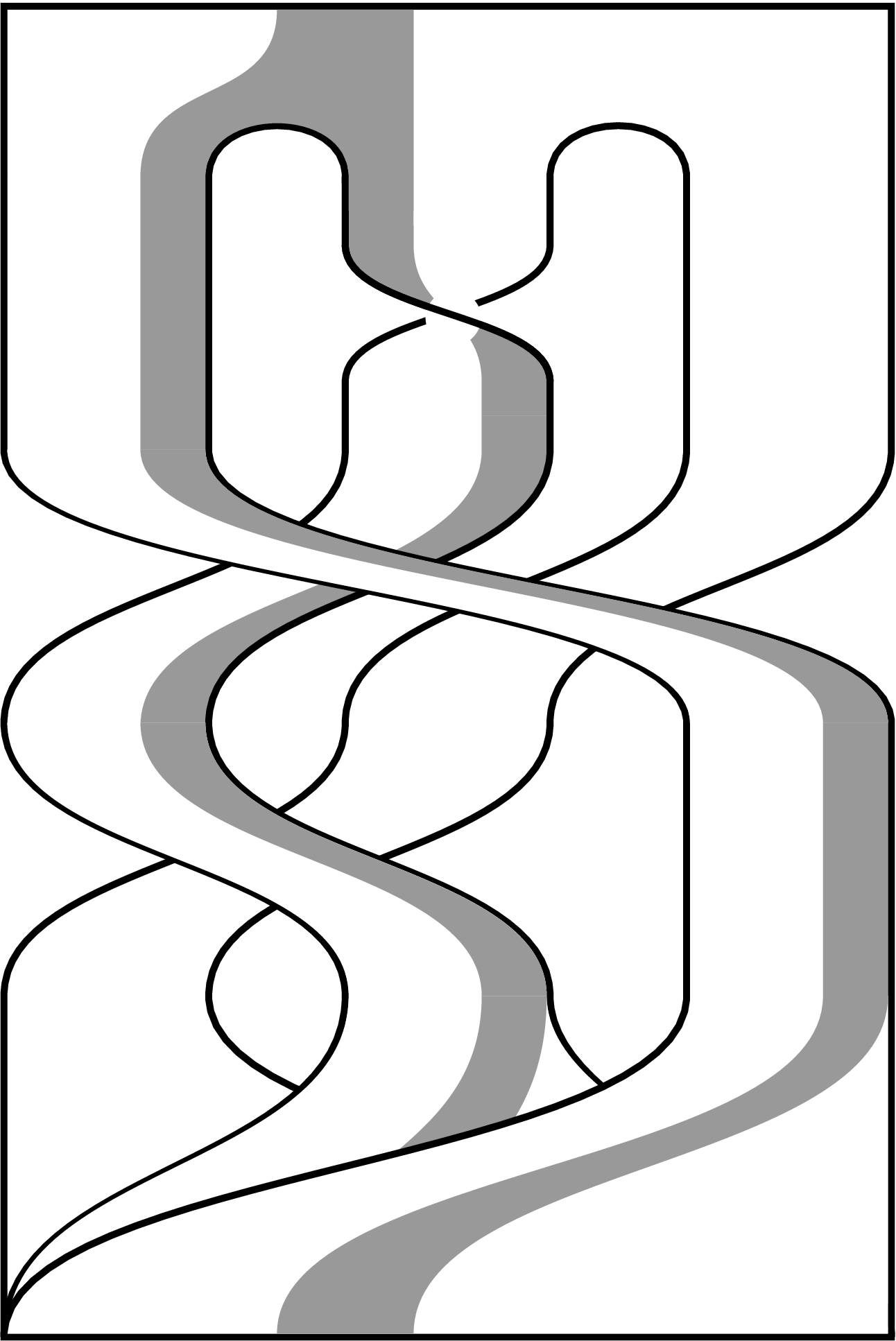} \\
    (a) $T_\mathcal{C}$ & (b) $T_\mathcal{A} \subset T_\mathcal{C}$
  \end{tabular}
  \vspace{-.5em}
  \caption{
    Template of $\mathcal{A}$ is a subtemplate of the template of
    $\mathcal{C}$.
  }
  \label{fig:rossler_A_subtemplate}
\end{figure}

The strip organization of $T_\mathcal{A}$ are the same of the two first
strips of $T_\mathcal{C}$. This means that $T_\mathcal{A}$ is a subtemplate
of $T_\mathcal{C}$: $T_\mathcal{A} \subset T_\mathcal{C}$. This is illustrated
on Fig.~\ref{fig:rossler_A_subtemplate} where we only display the mixers and
not the trivial part of the template that link the bottom to top on the left
side to have a clockwise flow as shown Fig.~\ref{fig:rossler_A_template}; this
is also the case for the remainder of this article. We will use
graphical representation of the templates and subtemplates because it details
the relation between template and subtemplate. This representation combined
with algebraical relations between linking matrices proves that a template is a
subtemplate of a template.

\subsubsection{Banded chaos}
\label{ssec:banded}

\begin{figure}[hbtp]
  \centering
  \begin{tabular}{cc}
    \includegraphics[width=.3\textwidth]{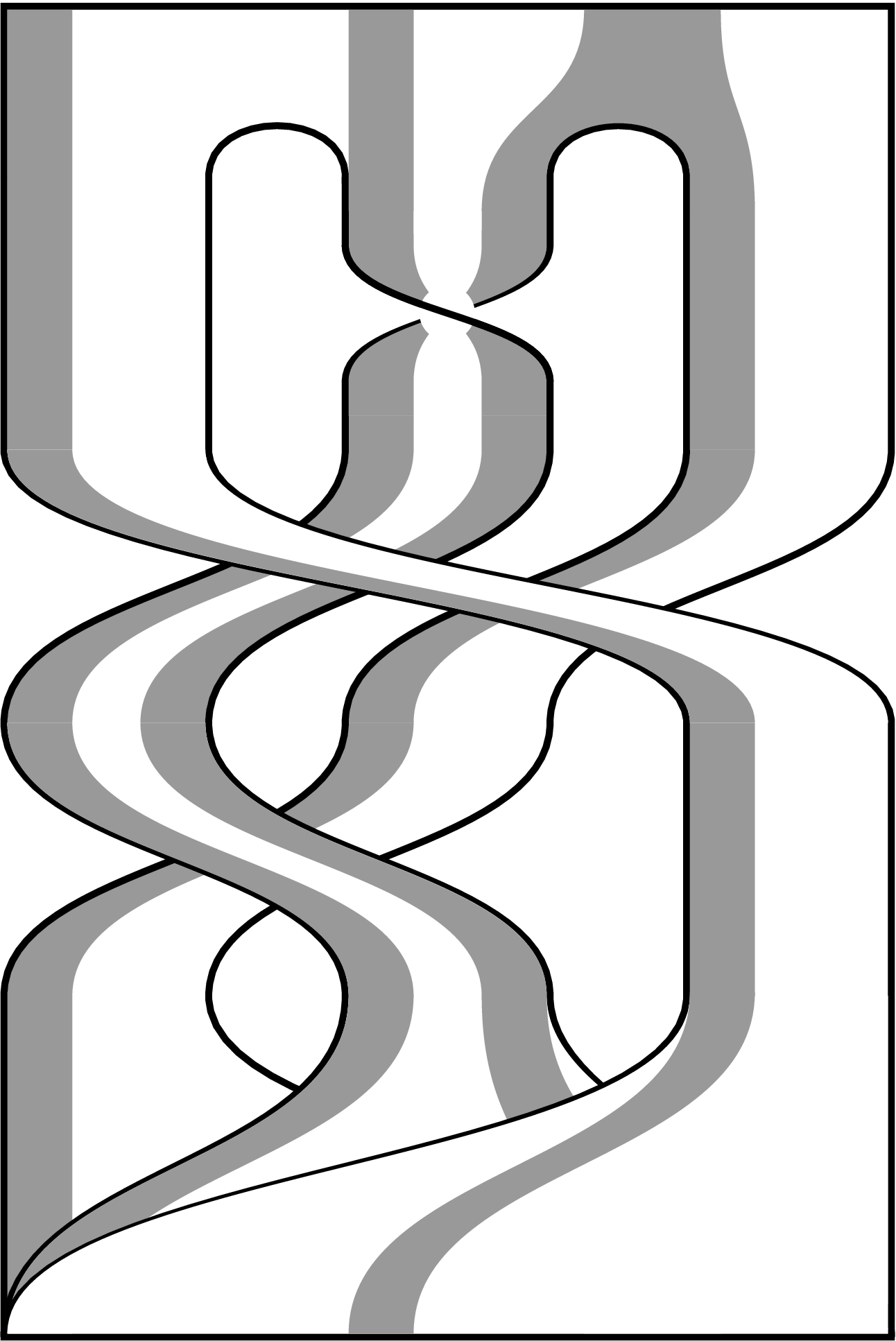} &
    \resizebox{.3\textwidth}{!}{\huge
        \begingroup%
        \makeatletter%
        \setlength{\unitlength}{371.83000488bp}%
        \makeatother%
        \begin{picture}(1,1.50888399)%
            \put(0,0){\includegraphics[width=\unitlength]{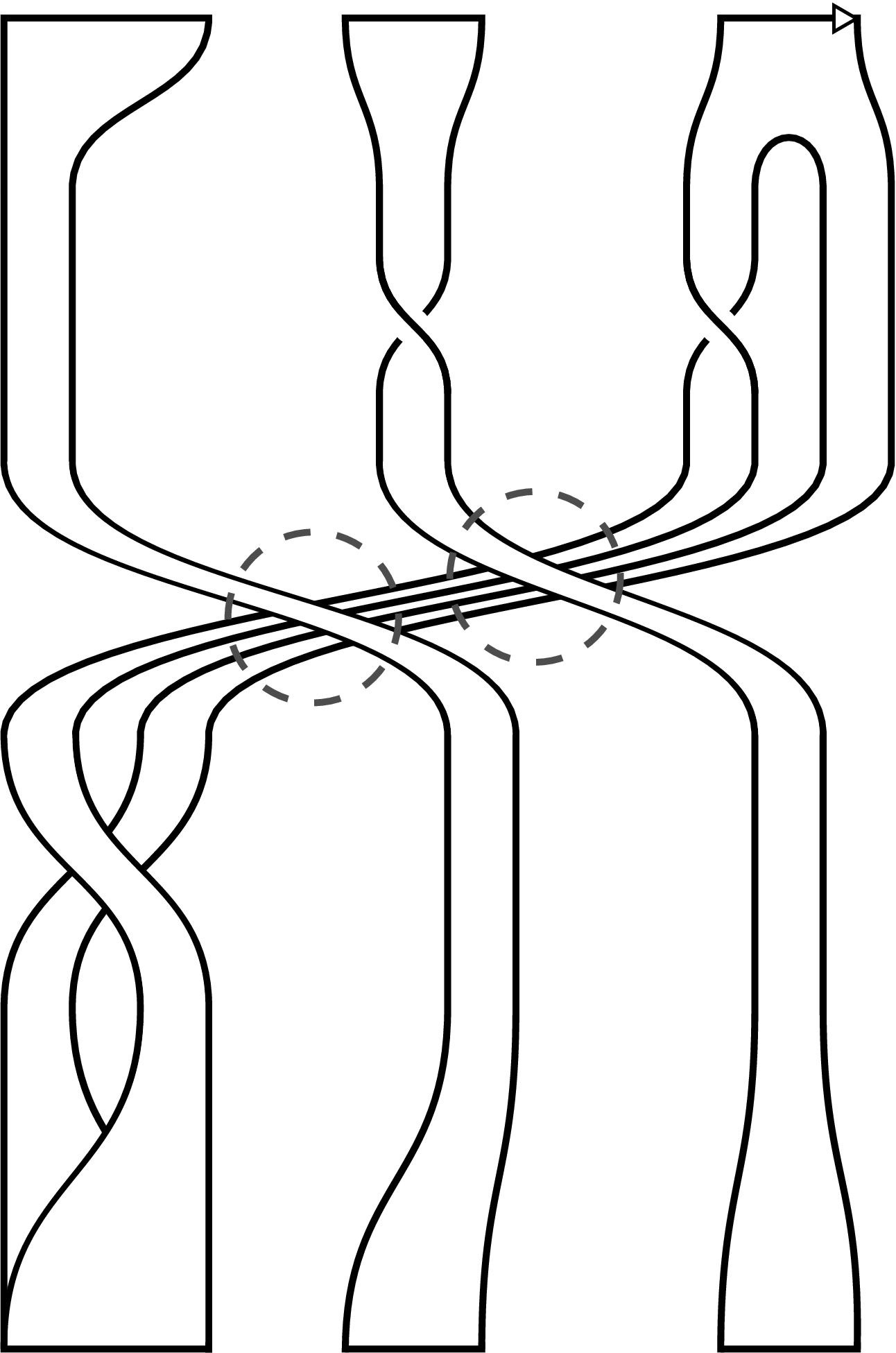}}%
            \put(0.11882467,1.41480716){\color[rgb]{0,0,0}\makebox(0,0)[b]{\smash{B}}}%
            \put(0.46188194,1.41480716){\color[rgb]{0,0,0}\makebox(0,0)[b]{\smash{C}}}%
            \put(0.88117426,1.41480716){\color[rgb]{0,0,0}\makebox(0,0)[b]{\smash{A}}}%
            \put(0.11882415,0.02351855){\color[rgb]{0,0,0}\makebox(0,0)[b]{\smash{B}}}%
            \put(0.46188194,0.02351855){\color[rgb]{0,0,0}\makebox(0,0)[b]{\smash{C}}}%
            \put(0.88117426,0.02351855){\color[rgb]{0,0,0}\makebox(0,0)[b]{\smash{A}}}%
        \end{picture}%
        \endgroup}\\
    (a) $T_\mathcal{B} \subset T_\mathcal{C}$ & (b) $T_\mathcal{B}$
  \end{tabular}
  \vspace{-.5em}
  \caption{
    (a) Template of attractor $\mathcal{B}$ is a subtemplate of the template
    of $\mathcal{C}$. (b) Template of attractor $\mathcal{B}$ shaped as a
    subtemplate of $\mathcal{C}$. The arrow over $A$ refers to the Poincar\'e
    section \eqref{eq:rossler_B_section} with the $\rho_n$ orientation.
  }
  \label{fig:rossler_B_subtemplate}
\end{figure}
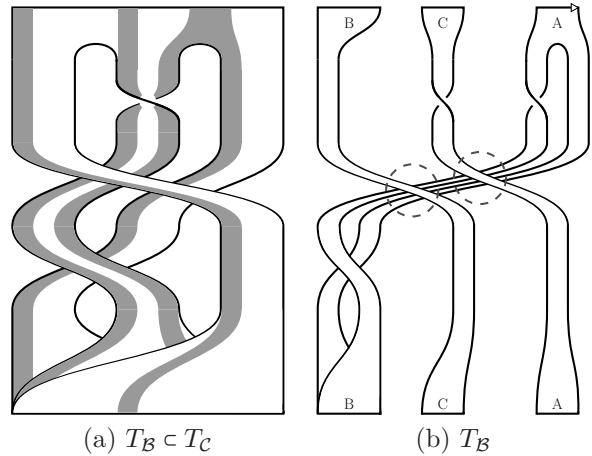

Attractors $\mathcal{B}$, $\mathcal{E}_1$ and $\mathcal{E}_2$ display
\emph{banded chaos} because they are composed by several strips, or bands,
with writhes.  We start with the template $T_\mathcal{B}$. We know that this
template can be considered with five negative torsions before a ``horseshoe''
mechanism \eqref{eq:torsion_mixer}. Thus we have to find a subtemplate that
goes through a ``horseshoe'' mechanism and have five negative torsions.
Letellier {\it et al.} \cite{letellier1999topological} underline the fact that
if a writhe is observed in an attractor bounded by a genus one torus, this is
equivalent to two torsions by isotopy; the sign of the torsions is the same of
the writhe one (see FIG.~3 and FIG.~4 of \cite{letellier1999topological} for
additional details). As a consequence, a subtemplate with $n$ portions induces
$n$ writhes that are equivalent to $2n$ torsions by isotopy; the sign of these
$2n$ torsions is the sign of the permutations of subtemplate portions.  We
propose to build such a subtemplate where
Fig.~\ref{fig:rossler_B_subtemplate}a is $T_\mathcal{B}$ as the subtemplate of
$T_\mathcal{C}$.

We propose to use algebraical relation between linking matrices to validate
subtemplates.  Fig.~\ref{fig:rossler_B_subtemplate}b is the template of
$\mathcal{B}$ shaped as a subtemplate of $\mathcal{C}$. When we establish the
template of $\mathcal{B}$, we chose to use the Poincar\'e section
\eqref{eq:rossler_B_section} that corresponds to the portion far from the inside
of the attractor; this portion is labeled $A$ on
Fig.~\ref{fig:rossler_B_subtemplate}b. We propose to concatenate its
components with respect to their relative order: from $A$ to $B$, then $B$ to $C$ and $C$ to $A$ that are
respectively: a mixer, a strip without torsion and a negative torsion.  We
finally consider the transformation by isotopy that does not have an impact on
the orientation of the strips. Thus we decide to concatenate the $2n$
torsions after the concatenation of the components. As illustrated
Fig.~\ref{fig:rossler_B_subtemplate}b, there is two negative permutations ($B$
to $C$ over $A$ to $B$ and $C$ to $A$ over $A$ to $B$) that are equivalent to
$2\times 2=4$ negative torsions (dot circles of
fig.~\ref{fig:rossler_B_subtemplate}b).
Consequently, we concatenate all these mixers and torsions
\begin{equation}
    \left[
      \begin{array}{C{1.3em}C{1.3em}}
        -1 & -1  \\
        -1 & 0
      \end{array} \right\rsem
+ \left[ -1 \right]
+ \left[ -2 \right]
+ \left[ -2 \right]
=
    \left[
      \begin{array}{C{1.3em}C{1.3em}}
        -6 & -6  \\
        -6 & -5
      \end{array} \right\rsem
      = T(\mathcal{B})
      \label{eq:BsubC}
\end{equation}
and obtain the linking matrix defining the template of $\mathcal{B}$.
Consequently $T_\mathcal{B}$ is a subtemplate of $T_\mathcal{C}$.

We now consider the two coexisting attractors $\mathcal{E}_1$ and
$\mathcal{E}_2$. They have a similar structure and coexist in the phase space
for distinct initial conditions. The mixer of $T_\mathcal{C}$ with three
branches has also a symmetric structure: the middle of the second strip of the
mixer $T(\mathcal{C})$ is a reflecting symmetry axis where the left side is
symmetric of the right side.  To build the $T_{\mathcal{E}_1}$ and
$T_{\mathcal{E}_2}$ as subtemplate of $T_\mathcal{C} $, we take this
symmetry into account.

\begin{figure}[htbp]
  \centering
  \begin{tabular}{ccc}
    \includegraphics[width=.3\textwidth]{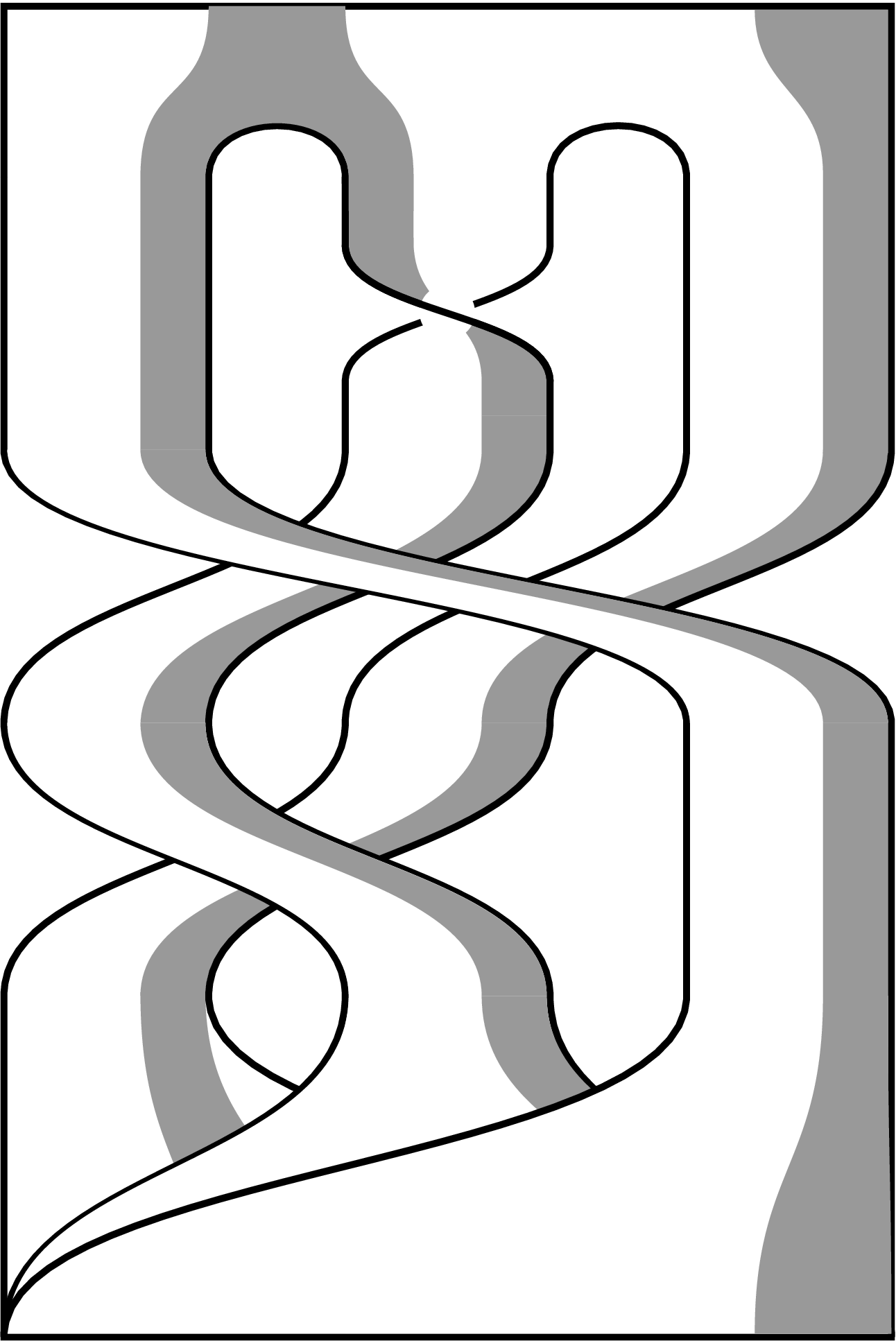} &
    \includegraphics[width=.3\textwidth]{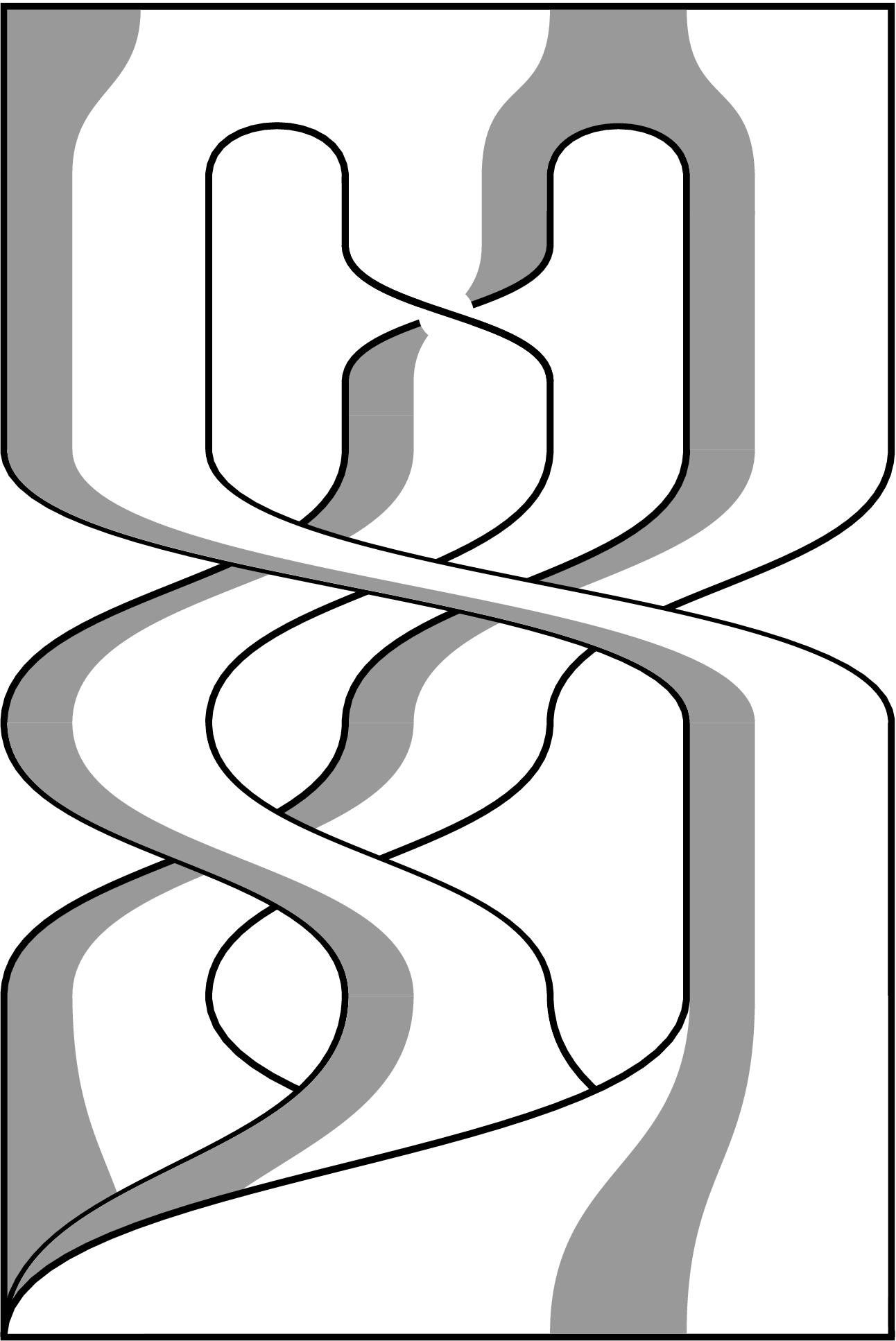} &
    \includegraphics[width=.3\textwidth]{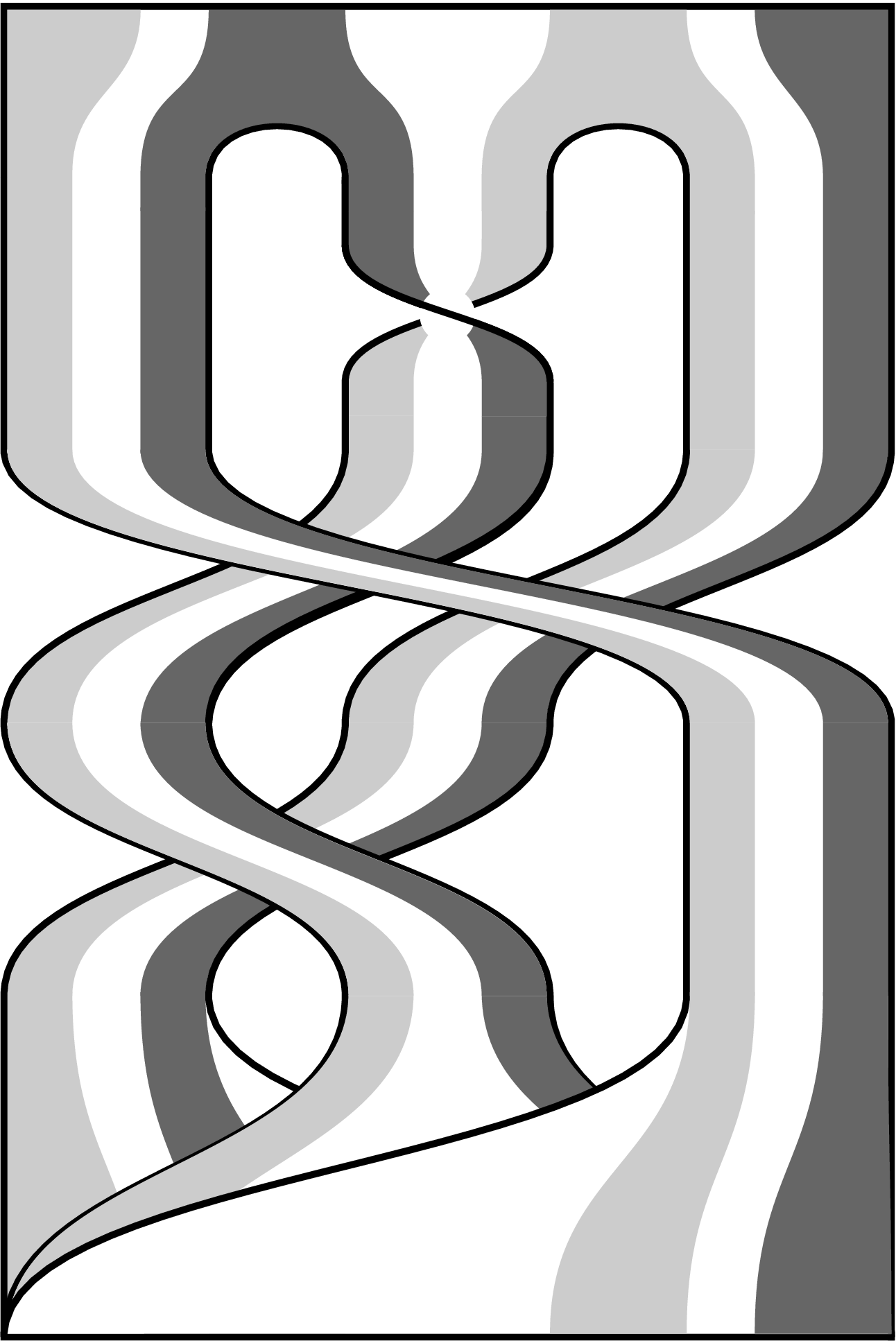}\\
    (a) $T_{\mathcal{E}_1} \subset T_\mathcal{C}$  &
    (b) $T_{\mathcal{E}_2} \subset T_\mathcal{C}$ &
    (c) $T_{\mathcal{E}_1}$ and $T_{\mathcal{E}_2}$ coexisting
  \end{tabular}
  \vspace{-.5em}
  \caption{
    Coexisting templates $T_{\mathcal{E}_1}$ and $T_{\mathcal{E}_2}$ are subtemplates of the template of
    $\mathcal{C}$.
  }
  \label{fig:rossler_E_subtemplate}
\end{figure}

We propose these subtemplates as illustrated
Fig.~\ref{fig:rossler_E_subtemplate}.  We compute the concatenation of
torsions and mixer that appears in these figures. For $T_{\mathcal{E}_1}$, we
have two parts: one is a mixer and the other is a strip without torsion.
These parts permute negatively once in a writhe; it algebraically corresponds
to a concatenation of two negative torsions. Thus, the linking matrix of such
a subtemplate (Fig.~\ref{fig:rossler_E_subtemplate}a) is
\begin{equation}
    \left[
      \begin{array}{C{1.3em}C{1.3em}}
        0 & -1  \\
        -1 & -1
      \end{array} \right\rsem
 + \left[ -2 \right]
=
    \left[
      \begin{array}{C{1.3em}C{1.3em}}
        -2 & -3  \\
        -3 & -3
      \end{array} \right\rsem
      = T(\mathcal{E}_1)\;.
\end{equation}
Similarly, the linking matrix of $T_{\mathcal{E}_2}$ as a subtemplate
(Fig.~\ref{fig:rossler_E_subtemplate}b) is
\begin{equation}
    \left[
      \begin{array}{C{1.3em}C{1.3em}}
        -1 & -1  \\
        -1 & 0
      \end{array} \right\rsem
 + \left[ -2 \right]
=
    \left[
      \begin{array}{C{1.3em}C{1.3em}}
        -3 & -3  \\
        -3 & -2
      \end{array} \right\rsem
      = T(\mathcal{E}_2)\;.
\end{equation}
These algebraical relations between template and subtemplate linking matrices
with Fig.~\ref{fig:rossler_E_subtemplate}
prove that $T_{\mathcal{E}_1} \subset T_\mathcal{C}$ and
$T_{\mathcal{E}_2} \subset T_\mathcal{C}$. Moreover, these two subtemplates
are symmetric one to the other by reflection and coexist in the template of
$\mathcal{C}$ (Fig.~\ref{fig:rossler_E_subtemplate}c).

\subsubsection{Concatenation of mixers}

\begin{figure}[hbtp]
  \centering
  \begin{tabular}{cc}
    \includegraphics[height=.3\textheight]{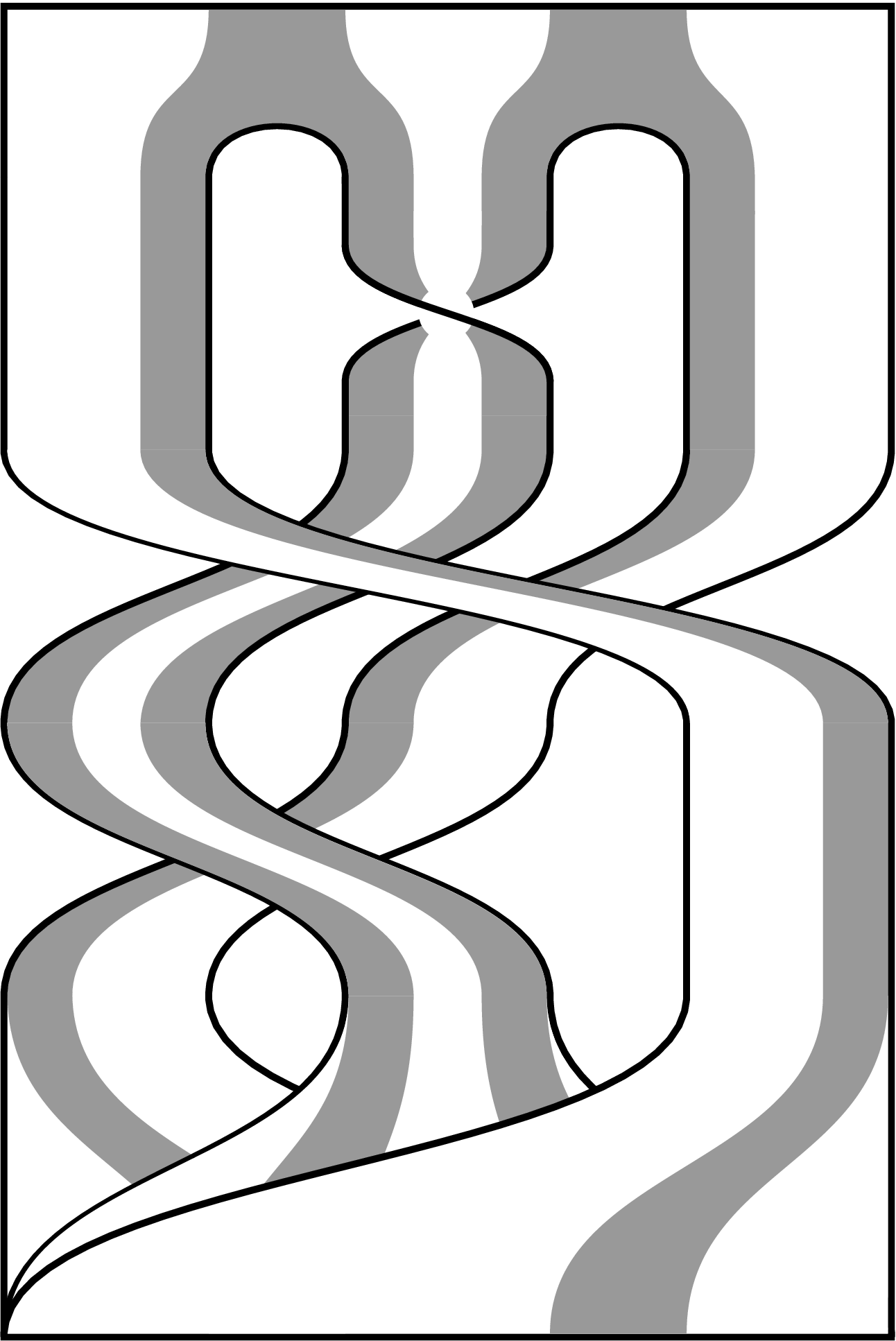} &
    \includegraphics[height=.3\textheight]{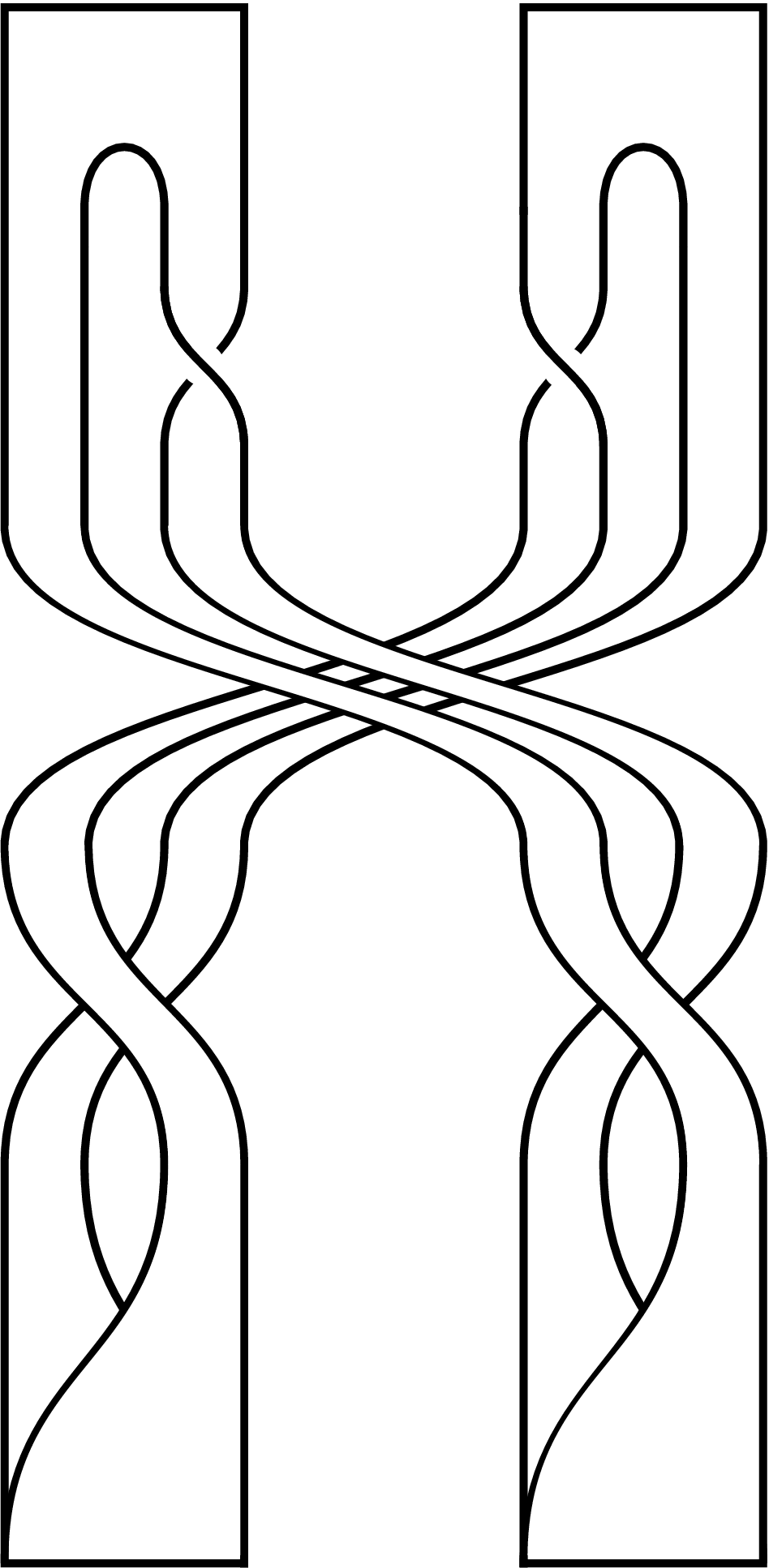}\\
    (a) $T_\mathcal{F} \subset T_\mathcal{C}$ & (b) $T_\mathcal{F}$
  \end{tabular}
  \vspace{-.5em}
  \caption{
    (a) The template of $\mathcal{F}$ is a subtemplate of the template of
    $\mathcal{C}$. (b) $T_\mathcal{F}$ with two mixers.
  }
  \label{fig:rossler_F_subtemplate}
\end{figure}

We now consider $T_\mathcal{F}$, the template of $\mathcal{F}$, made of four
strips. In a previous paper \cite{rosalie2015systematic}, we demonstrate that
the concatenation of two mixers is a mixer and its number of strips is the
product of the number of strips of each mixer.
Thus, the concatenation of two mixers made of two branches is a mixer with four
branches. Our hypothesis is that the four strips of $T_\mathcal{F}$ are the
result of this process. To validate it, we draw its subtemplate by splitting
the template of $\mathcal{C}$ into the two symmetric part containing a mixer;
we obtain the Fig.~\ref{fig:rossler_F_subtemplate}a.

As we do previously, we decompose this subtemplate
(Fig.~\ref{fig:rossler_F_subtemplate}b) in parts: the two parts contain a
mixer and these parts permute negatively once. Thus, we concatenate a mixer
before a mixer and two negative torsions (cf. \ref{ssec:banded}).  The first
concatenation gives a mixer defined by a linking matrix; the algebraical
relation necessary to obtain this matrix are detailed in
\cite{rosalie2015systematic}.  The linking matrix of $T_\mathcal{F}$ as a
subtemplate (Fig.~\ref{fig:rossler_F_subtemplate}a) is:
{\small
  \begin{equation}
    \begin{array}{l}
    \left[ \begin{matrix} -1 & -1 \\ -1 & 0
  \end{matrix}\right\rsem +
\left[ \begin{matrix} 0 & -1 \\ -1 & -1 \end{matrix}\right\rsem + [-2] \\
   \quad =\left[\left|
    \begin{array}{C{1.3em}C{1.3em}C{1.3em}C{1.3em}}
      -1 & -1 & -1 & -1 \\
      -1 & -1 & -1 & -1 \\
      -1 & -1 & 0 & 0 \\
      -1 & -1 & 0 & 0
    \end{array}\right|
    +
    \left|
    \begin{array}{C{1.3em}C{1.3em}C{1.3em}C{1.3em}}
      0 & 0 & 0 & 0 \\
      0 & 0 & 0 & 1 \\
      0 & 0 & 0 & 0 \\
      0 & 1 & 0 & 0
  \end{array}\right|
  +
  \left|
    \begin{array}{C{1.3em}C{1.3em}C{1.3em}C{1.3em}}
    -1 & -1 & -1 & -1 \\
    -1 & 0 & 0 & -1 \\
    -1 & 0 & 0 & -1 \\
    -1 & -1 & -1 & -1
  \end{array}\right| \right\rsem + [-2] \\
  \quad =
  \left[
    \begin{array}{C{1.3em}C{1.3em}C{1.3em}C{1.3em}}
      -2 & -2 & -2 & -2  \\
      -2 & -1 & -1 & -1  \\
      -2 & -1 & 0 & -1  \\
      -2 & -1 & -1 & -1
    \end{array}\right\rsem + [-2] 
     = 
  \left[
    \begin{array}{C{1.3em}C{1.3em}C{1.3em}C{1.3em}}
      -4 & -4 & -4 & -4  \\
      -4 & -3 & -3 & -3  \\
      -4 & -3 & -2 & -3  \\
      -4 & -3 & -3 & -3
    \end{array}\right\rsem =
     T(\mathcal{F})
   \end{array}
  \end{equation}
  }
This algebraical relation between template and subtemplate linking matrices
associated to Fig.~\ref{fig:rossler_F_subtemplate} prove that $T_\mathcal{F}
\subset T_\mathcal{C}$.

\begin{figure}[hbtp]
  \centering
  \includegraphics[height=.3\textheight]{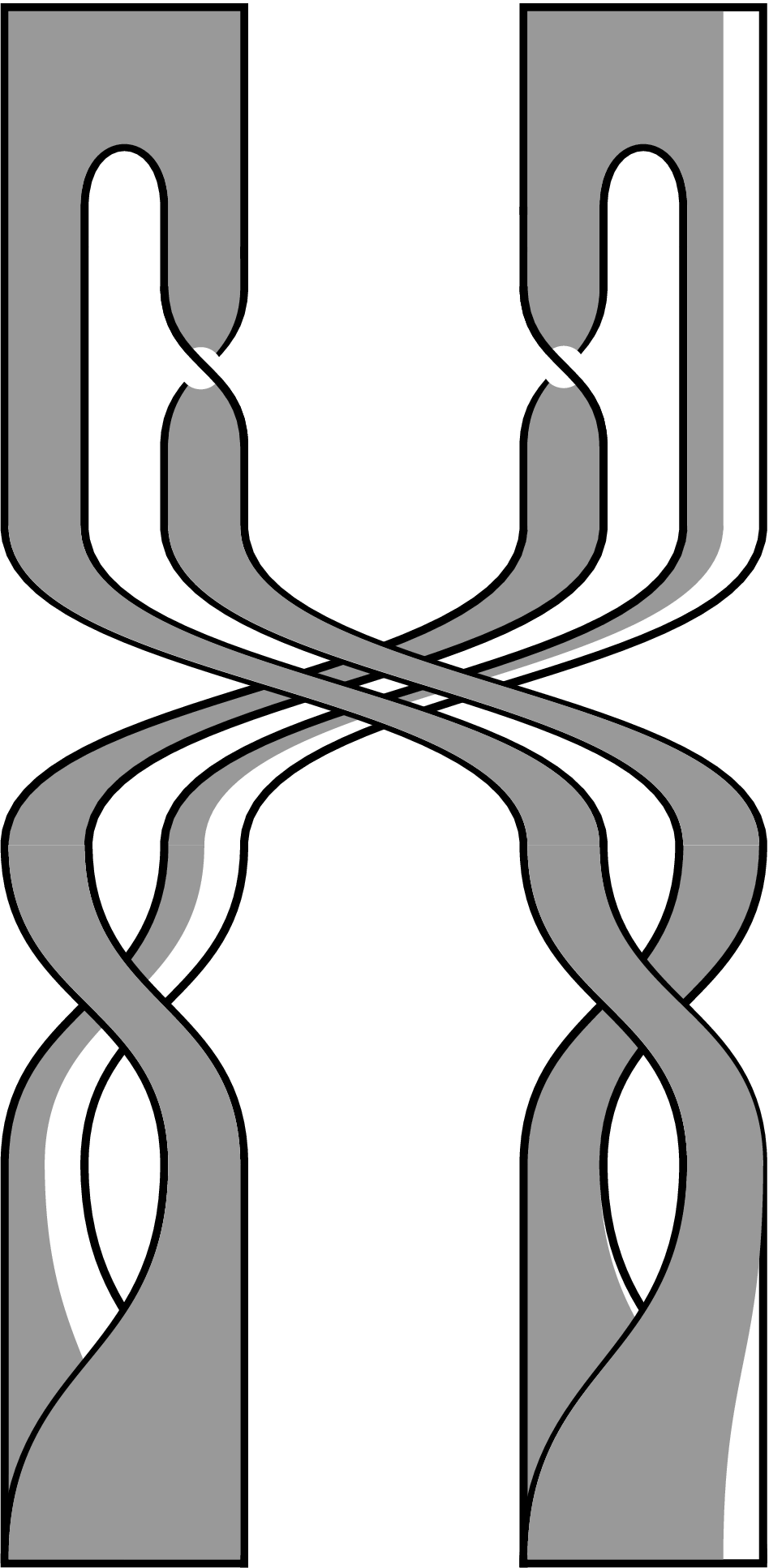}
  \caption{
    Template of $\mathcal{D}$ is a subtemplate of the template of
    $\mathcal{F}$ and consequently, it is also a subtemlate of the template of
    $\mathcal{C}$.
  }
  \label{fig:rossler_D_subtemplate}
\end{figure}

We now consider the attractor $\mathcal{D}$, we have $T_\mathcal{D} \subset T_\mathcal{F}$
directly from their mixers
\begin{equation}
\left[
    \begin{array}{C{1.3em}C{1.3em}C{1.3em}}
      -4 & -4 & -4 \\
      -4 & -3 & -3 \\
      -4 & -3 & -2
    \end{array}
  \right\rsem \subset
 \left[
    \begin{array}{C{1.3em}C{1.3em}C{1.3em}C{1.3em}}
      -4 & -4 & -4 & -4  \\
      -4 & -3 & -3 & -3  \\
      -4 & -3 & -2 & -3  \\
      -4 & -3 & -3 & -3
    \end{array}
  \right\rsem\;;
\end{equation}
this is illustrated on Fig.~\ref{fig:rossler_D_subtemplate}. We previously
obtain that $T_\mathcal{F} \subset T_\mathcal{C}$, thus we prove that $T_\mathcal{D} \subset
T_\mathcal{C}$.

Consequently, we prove that the six templates of attractors $\mathcal{A}$,
$\mathcal{B}$, $\mathcal{D}$, $\mathcal{E}_1$, $\mathcal{E}_2$ and
$\mathcal{F}$ are subtemplate of the template of the attractor $\mathcal{C}$;
it is a template with six subtemplates. We remind that $\mathcal{G}$ and
$\mathcal{C}$ have the same template ($T_\mathcal{G}=T_\mathcal{C}$).

\section{A template for the whole bifurcation diagram}

We obtain a unique template containing the eight templates of attractors. We
now consider the whole bifurcation diagram
(Fig.~\ref{fig:rossler_bifurcation}) and not only specific attractors. In this
section, we will show that the template of $\mathcal{C}$ contains all
attractors templates for any parameter value take from this bifurcation
diagram. We use the Poincar\'e section \eqref{eq:rossler_X_section} and build
return maps using $y_n$ for $\alpha \in ]-2;1.8[$ when an attractor is
solution. We associate a symbolic dynamic with the three symbols ``$0$'',
``$1$'', ``$2$'' of $T_\mathcal{C}$.  Note that Barrio {\it et al.}
\cite{barrio2012topological} also use this process to study return maps of a
R\"ossler system from a Lyapunov diagram. The authors display return maps with
superstability curve and coexisting stable points.  Here we prefer collect
extrema points to make a partition of the bifurcation diagram.

\begin{figure}[htpb]
  \centering
  \includegraphics[width=.7\textwidth]{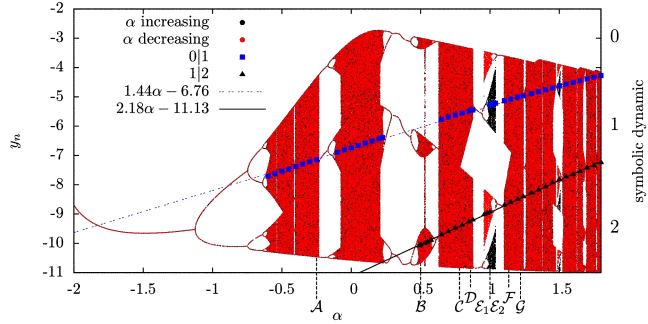}
  \caption{
    Partition of the bifurcation diagram when $\alpha$ varies build using
    first return maps on $y_n$ of the Poincar\'e section
    \eqref{eq:rossler_X_section}. This partition give a symbolic dynamic with
    three symbols ``$0$'', ``$1$'', ``$2$'' depending on $\alpha$.
  }
  \label{fig:rossler_bifurcation_symbolic}
\end{figure}

In the diagram Fig.~\ref{fig:rossler_bifurcation_symbolic}, we indicate the
values of $y_n$ splitting the return maps into two or three parts. We remind
the reader that we orientate application from the inside to the outside of the
attractor. Thus, the branches are labelled with symbol number increasing while
$y_n$ decrease. Fig.~\ref{fig:rossler_bifurcation_symbolic} reveals that the
separator values are linear to $\alpha$. We note $y_{0|1}(\alpha)$
the value of $y_n$ that split branches ``$0$'' and ``$1$'' and
$y_{1|2}(\alpha)$ the value of $y_n$ that split branches ``$1$'' and ``$2$''. A
linear regression gives
\begin{equation}
  \begin{array}{l}
    y_{0|1}(\alpha) = 1.43638 \alpha -6.76016 \\
    y_{1|2}(\alpha) = 2.18237 \alpha -11.1289 \;.
  \end{array}
  \label{eq:partition}
\end{equation}
Up to this point, if there is an attractor solution, its orbits can be encoded
with the symbols depending on the previous equations. This also requires the
use of the Poincar\'e section \eqref{eq:rossler_X_section}.

For a given range of a bifurcation parameters ($\alpha \in ]-2;1.8[$), the
parameters of the R\"ossler system depends on $\alpha$: $a(\alpha)$,
$b(\alpha)$ and $c(\alpha)$. The fixed points depend on the parameters and
the Poincar\'e section depend on the fixed points. Thus, we obtain a
Poincar\'e section and its partition \eqref{eq:partition} depending on
$\alpha$ while the template is define by the linking matrix
\begin{equation}
\left[
    \begin{array}{C{1.3em}C{1.3em}C{1.3em}}
       0 & -1 & -1  \\
      -1 & -1 & -1  \\
      -1 & -1 &  0
    \end{array}
  \right\rsem\;.
\end{equation}

The main result is that the topological characterization of chaotic attractors
can be extended as a description of various attractors whose parameters come
from one bifurcation diagram. In this bifurcation diagram, we show that there
are regimes where the chaotic mechanisms are topologically equivalent
($T_\mathcal{C} = T_\mathcal{G}$), symmetric ($T_{\mathcal{E}_1}$ and
$T_{\mathcal{E}_2}$) and they are a subset of the same chaotic mechanism.  The
point is that our work can help to understand the complex structure of
attractors considering them as subtemplates of their neighbors (in term of
bifurcation parameter). This also enlarge the possibility to use the
topological characterization to describe more than an attractor, but an entire
bifurcation diagram.

\section{Conclusion}

In this paper we study eight attractors of the R\"ossler system. The
parameters values of these attractors come from a bifurcation diagram that
exhibits various dynamics such as coexisting attractors. For each attractor we
apply the topological characterization method that give us a template of the
attractor. These templates detail the chaotic mechanism and these are only
made of stretching and folding mechanism followed by a squeezing mechanism
with two, three and four strips.

The second part of this paper is dedicated to the proof that the eights
templates are subtemplates of a unique template: the template of
$\mathcal{C}$. The main result here is that a template is no longer a tool to
describe one attractor but also a set of neighbours attractors (in the
parameter space). Thus for a bifurcation diagram, we build a partition using
symbols of $T_\mathcal{C}$.  This partition over the whole diagram give a
global structure of attractors for a range of parameters.

This better understanding of the structure of bifurcation diagram can help
researchers that want to explore the behaviour of their system, especially if
it exhibits chaotic properties. For instance Matthey \textit{et al.}
\cite{matthey2008experimental} design a robot using coupled R\"ossler
oscillators to simulate its locomotion.  A similar theoretical analysis of
their system can provide various set of parameters with specific chaotic
properties that might induce new locomotion pattern. A partitioned bifurcation
diagram details the various non equivalent dynamical behavior of the system to
find them.

This work on templates of attractors from a unique bifurcation diagram is a
first step that can lead to a description of manifolds using templates. It is
also a new way to classify templates grouping them as subtemplates and not
claiming that it exists six new attractors for the R\"ossler system. To
conclude, this work permits to apply the topological characterization method
to a set of attractors from a bifurcation diagram. The partition of a
bifurcation diagram associated with a unique template is a new tool to
describe the global dynamical properties of a system while a parameter is
varied. In future works, we will apply this method on attractors bounded by
higher genus torus to highlight how symmetry breaking or template number of
branches are related in a bifurcation diagram.

\end{document}